\documentclass[aps,pra,nobalancelastpage,superscriptaddress,twocolumn]{revtex4-2}
\usepackage{mathtools}
\DeclarePairedDelimiter{\ceil}{\lceil}{\rceil}
\usepackage{tensor, scalerel, comment}
\usepackage{bbm}
\usepackage[usenames,dvipsnames,x11names]{xcolor}
\usepackage{amsmath, amssymb, amsfonts, amsthm}
\usepackage{array}

\usepackage{physics}
\usepackage{hyperref}
\usepackage{pgfplots}
\usepackage{tikz}
\usepackage{tikz-network}
\usetikzlibrary{decorations.pathreplacing,calligraphy}
\usepackage{pgf} 
\usepackage{MnSymbol,wasysym}
\usepackage{dsfont}
\usepackage{empheq}
\usepackage{xcolor}
\definecolor{myred}{RGB}{232,102,102}
\definecolor{myblue}{RGB}{101,147,245}
\definecolor{mygreyblue}{RGB}
{166,218,149}
\definecolor{mygreyred}{RGB}
{232,102,102}
\definecolor{myvioletc}{RGB}{45,130,60}
\definecolor{mygreen}{RGB}{187,187,255}
\definecolor{green}{RGB}{76,187,23}
\definecolor{myorange}{RGB}
{255,145,33}
\definecolor{OliveGreen}{RGB}{85,107,47}
\definecolor{NavyBlue}{RGB}{0,0,128}
\definecolor{myviolet}{RGB}{210,145,178}
\definecolor{mygrey1}{RGB}{220,220,220}
\definecolor{mygrey2}{RGB}{211,211,211}
\definecolor{mygrey3}{RGB}{192,192,192}
\definecolor{mygrey4}{RGB}{169,169,169}

\tikzset{
	Ufolded/.pic = {
		\draw[thick] (-.25,0) -- (0.25,.5);
		\draw[thick] (-0.25,.5) -- (0.25,0);
		\draw[thick, fill=myorange, rounded corners=2pt] (-0.125,0.375) rectangle (.125,0.125);
		\draw[thick] (0,0.325) -- (.075,0.325) -- (.075,0.25);
}}
\tikzset{
	Ufoldedprime/.pic = {
		\draw[thick] (-.25,0) -- (0.25,.5);
		\draw[thick] (-0.25,.5) -- (0.25,0);
		\draw[thick, fill=myorange, rounded corners=2pt] (-0.125,0.375) rectangle (.125,0.125);
		\draw[thick] (0,0.325) -- (-.075,0.325) -- (-.075,0.25);
}}
\tikzset{
	rectp/.pic = {
		\draw[thick,fill=white] (-0.05 ,-0.05) rectangle (+0.05,0.05);
		\node at (0.07 ,.02) {$'$};
}}\tikzset{
	rectb/.pic = {
		\draw[thick,fill=black] (-0.05 ,-0.05) rectangle (+0.05,0.05);
}}\tikzset{
	circp/.pic = {
		\draw[thick,fill=white] (0,0) circle (.07);	
		\node at (0.09 ,.05) {$'$};
}}\tikzset{
	circpp/.pic = {
		\draw[thick,fill=white] (0,0) circle (.07);	
		\node at (-0.09 ,.05) {$''$};
}}\tikzset{
	circ2/.pic = {
		\draw[thick,fill=white] (0,0) circle (.07);
	}
}
\tikzset{
	circ2b/.pic = {
		\draw[thick,fill=black] (0,0) circle (.07);
	}
}
\tikzset{
	rec2/.pic = {
		\draw[thick,fill=white] (-.06,.06) rectangle (.06,-.06);
	}
}
\tikzset{rec2b/.pic = {
		\draw[thick,fill=black] (-.06,.06) rectangle (.06,-.06);
	}
}

\tikzset{
	Uffolded/.pic = {
		\draw[thick] (-.25,0) -- (0.25,.5);
		\draw[thick] (-0.25,.5) -- (0.25,0);
		\draw[thick, fill=myvioletc, rounded corners=2pt] (-0.125,0.375) rectangle (.125,0.125);
}}
\tikzset{
	Ufoldeddag/.pic = {
		\draw[thick] (-.25,0) -- (0.25,.5);
		\draw[thick] (-0.25,.5) -- (0.25,0);
		\draw[thick, fill=blue, rounded corners=2pt] (-0.125,0.375) rectangle (.125,0.125);
}}
\tikzset{
	Ufoldeddagrot/.pic = {
		\draw[thick] (-.25,0) -- (0.25,.5);
		\draw[thick] (-0.25,.5) -- (0.25,0);
		\draw[thick, fill=blue, rounded corners=2pt] (-0.125,0.375) rectangle (.125,0.125);
		\draw[thick] (0,.5-0.325) -- (-.075,.5-0.325) -- (-.075,.5-0.25);
}}\tikzset{
	Ufoldeddagrotprime/.pic = {
		\draw[thick] (-.25,0) -- (0.25,.5);
		\draw[thick] (-0.25,.5) -- (0.25,0);
		\draw[thick, fill=blue, rounded corners=2pt] (-0.125,0.375) rectangle (.125,0.125);
		\draw[thick] (0,0.325) -- (-.075,0.325) -- (-.075,0.25);
}}
\tikzset{
	U/.pic = {
		\draw[thick] (-.25,0) -- (0.25,.5);
		\draw[thick] (-0.25,.5) -- (0.25,0);
		\draw[thick, fill=red, rounded corners=2pt] (-0.125,0.375) rectangle (.125,0.125);
		\draw[thick] (0,0.325) -- (.075,0.325) -- (.075,0.25);
}}
\tikzset{
	Udag/.pic = {
		\draw[thick] (-.25,0) -- (0.25,.5);
		\draw[thick] (-0.25,.5) -- (0.25,0);
		\draw[thick, fill=myblue, rounded corners=2pt] (-0.125,0.375) rectangle (.125,0.125);
		\draw[thick] (0,0.5-0.325) -- (-.075,0.5-0.325) -- (-.075,0.5-0.25);}}
\tikzset{
	U1/.pic = {
		\draw[thick] (-.25,0) -- (0.25,.5);
		\draw[thick] (-0.25,.5) -- (0.25,0);
		\draw[thick, fill=red, rounded corners=2pt] (-0.125,0.375) rectangle (.125,0.125);
		\draw[thick] (0,0.5-0.325) -- (-.075,0.5-0.325) -- (-.075,0.5-0.25);}}
\tikzset{
	Udag1/.pic = {
		\draw[thick] (-.25,0) -- (0.25,.5);
		\draw[thick] (-0.25,.5) -- (0.25,0);
		\draw[thick, fill=myblue, rounded corners=2pt] (-0.125,0.375) rectangle (.125,0.125);
		\draw[thick] (0,0.325) -- (-.075,0.325) -- (-.075,0.25);}}
\tikzset{
	Udag2/.pic = {
		\draw[thick] (-.25,0) -- (0.25,.5);
		\draw[thick] (-0.25,.5) -- (0.25,0);
		\draw[thick, fill=myblue, rounded corners=2pt] (-0.125,0.375) rectangle (.125,0.125);
		\draw[thick] (0,.5-0.325) -- (.075,.5-0.325) -- (.075,.5-0.25);}}
\tikzset{
	Ufoldedupsidedown/.pic = {
		\draw[thick] (-.25,0) -- (0.25,.5);
		\draw[thick] (-0.25,.5) -- (0.25,0);
		\draw[thick, fill=myorange, rounded corners=2pt] (-0.125,0.375) rectangle (.125,0.125);
		\draw[thick] (0,0.5-0.325) -- (-.075,0.5-0.325) -- (-.075,0.5-0.25);
}}\tikzset{
	Udagrotated1/.pic = {
		\draw[thick] (-.25,0) -- (0.25,.5);
		\draw[thick] (-0.25,.5) -- (0.25,0);
		\draw[thick, fill=mygreen, rounded corners=2pt] (-0.125,0.375) rectangle (.125,0.125);
		\draw[thick] (0,0.325) -- (-.075,0.325) -- (-.075,0.25);
}}
\tikzset{
	Udagrotated2/.pic = {
		\draw[thick] (-.25,0) -- (0.25,.5);
		\draw[thick] (-0.25,.5) -- (0.25,0);
		\draw[thick, fill=mygreen, rounded corners=2pt] (-0.125,0.375) rectangle (.125,0.125);
		\draw[thick] (0,.5-0.325) -- (.075,.5-0.325) -- (.075,.5-0.25);
}}
\tikzset{
	U_dag/.pic = {
		\draw[thick] (-.25,0) -- (0.25,.5);
		\draw[thick] (-0.25,.5) -- (0.25,0);
		\draw[thick, fill=myblue, rounded corners=2pt] (-0.125,0.375) rectangle (.125,0.125);
		\draw[thick] (0,0.325) -- (.075,0.325) -- (.075,0.25);
}}

\tikzset{
	U_dagfolded/.pic = {
		\draw[thick] (-.25,0) -- (0.25,.5);
		\draw[thick] (-0.25,.5) -- (0.25,0);
		\draw[thick, fill=mygreen, rounded corners=2pt] (-0.125,0.375) rectangle (.125,0.125);
		\draw[thick] (0,0.325) -- (.075,0.325) -- (.075,0.25);
}}
\tikzset{
	Uvertical/.pic = {		\draw[thick] (0,-.375) -- (0,.375);
		\draw[thick] (.375,0) -- (-.375,0);
		\draw[thick, fill=mygreen, rounded corners=2pt,rotate around={45:(0,0)}] (-0.125,-0.125) rectangle (0.125,0.125);
		\draw[thick] (-.075,0) -- (0,0.075) -- (.075,0);
}}
\tikzset{
	Uvertical2/.pic = {		\draw[thick] (0,-.375) -- (0,.375);
		\draw[thick] (.375,0) -- (-.375,0);
		\draw[thick, fill=myred, rounded corners=2pt,rotate around={45:(0,0)}] (-0.125,-0.125) rectangle (0.125,0.125);
		\draw[thick] (0,-.075) -- (0.075,0) -- (0,.075);
}}
\tikzset{
	MPSstate/.pic = {
		\draw[very thick] (-0.5,0.25) -- (0.5,0.25);
		\draw[ thick, fill=myred, rounded corners=1pt] (-.25,0.5) rectangle (0.25,0.2);
		\draw[thick] (.05,0.125+.25+.05)-- (0.125+.05,0.125+.25+.05) -- (0.125+.05,.25+.05);	
		\draw[thick] (-.125-.25,0.125+.5)--(-.25,0.5);
		
		\draw[thick] (.125+.25,0.125+.5)--(.25,0.5);
}}
\tikzset{
	MPSstatefolded/.pic = {
		\draw[very thick] (-0.5,0.25) -- (0.5,0.25);
		\draw[ thick, fill=red, rounded corners=1pt] (-.25,0.5) rectangle (0.25,0.2);
		\draw[thick] (.05,0.125+.25+.05)-- (0.125+.05,0.125+.25+.05) -- (0.125+.05,.25+.05);	
		\draw[thick] (-.125-.25,0.125+.5)--(-.25,0.5);
		
		\draw[thick] (.125+.25,0.125+.5)--(.25,0.5);
}}
\tikzset{
	MPSstatefoldeddag/.pic = {
		\draw[very thick] (-0.5,0.25) -- (0.5,0.25);	
		\draw[thick] (-.125-.25,-0.125)--(-.25,0);
		\draw[thick] (.125+.25,-0.125)--(.25,0);
		\draw[ thick, fill=myblue, rounded corners=1pt] (-.25,0.5-.25) rectangle (0.25,0.2-.25);
		\draw[thick] (.05,.5-0.125-.25-.1)-- (0.125+.05,.5-0.125-.25-.1) -- (0.125+.05,.5-.25-.1);	
}}\tikzset{
	connectionMPSsqrd/.pic = {
		\draw (-.25,0)--(0,-.25)--(0.5,.25)--(0.25,.5);
		\draw[fill=black] (.25-.1,0+.1)--(.25+.1,0+.1)--(.25+.1,0-.1)--(.25-.1,0+.1);
}}
\tikzset{
	MPSsqrd/.pic = {
		\draw[fill=black] (0,.1414)--(0,-.1414)--(.1414,0);
}}
\tikzset{
	MPSsqrdstate/.pic = {
		\draw (+.14,0)--(0.0,0+.0)--(0,+.14);}}
\tikzset{
	Phi4folded/.pic = {
		\draw[thick] (-.25,0) -- (0.25,.5);
		\draw[thick] (-0.25,.5) -- (0.25,0);
		\draw[thick, fill=myblue, rounded corners=2pt] (-0.125,0.375) rectangle (.125,0.125);
}}

\tikzset{
	cross/.pic = {
		\draw[thick, fill=white] (0,0) circle (0.075);

	}
}

\tikzset{
	MPSpairstate1f/.pic = {		
		\draw[thick] (-.25,0.5) -- (0.0,.25)-- (0.25,.5);
		\draw[thick,fill=orange] (0.0,.25) circle(1.5pt);
}}
\tikzset{
	MPSpairstate/.pic = {		
		\draw[thick] (-.25,0.5) -- (0.0,.25)-- (0.25,.5);
		\draw[thick,fill=red] (0.0,.25) circle(1.5pt);
}}
\tikzset{
	MPSpairstatesolv/.pic = {		
		\draw[thick] (-.25,0.5) -- (0.0,.25)-- (0.25,.5);
}}
\tikzset{
	MPSpairstate1fprime/.pic = {		
		\draw[thick] (-.25,0.5) -- (0.0,.25)-- (0.25,.5);
		\draw[thick,fill=blue] (0.0,.25) circle(1.5pt);
}}
\tikzset{
	MPSpairstatef/.pic = {		
		\draw[thick] (-.25,0.5) -- (0.0,.25)-- (0.25,.5);
		\draw[thick,fill=myvioletc] (0.0,.25) circle(1.5pt);
}}
\tikzset{
	pairidentity/.pic = {		
		\draw[thick] (-.25,0.5) -- (0.0,.25)-- (0.25,.5);
}}
\tikzset{
	MPSpairstatefolded/.pic = {		
		\draw[thick] (-.25,0.5) -- (0.0,.25)-- (0.25,.5);
		\draw[thick,fill=orange] (0.0,.25) circle(1.5pt);
}}
\tikzset{
	MPSpairstatefoldeddag/.pic = {		
		\draw[thick] (-.25,0) -- (0.0,.25)-- (0.25,0);
		\draw[thick,fill=blue] (0.0,.25) circle(1.5pt);
}}
\tikzset{
	MPSpairstatevertical/.pic = {		
		\draw[thick] (0,-.25)--(0,0) -- (.25,0);
		\draw[thick,fill=black] (0,0) circle(1.5pt);
}}
\tikzset{
	pairidentityvertical/.pic = {		
		\draw[thick] (-.25,0.5) -- (0,.25)-- (-0.25,0);
}}

\tikzset{
	MPSpairstateupsidedown/.pic = {		
		\draw[thick] (-.25,0) -- (0.0,.25)-- (0.25,0);
		\draw[thick,fill=myblue] (0.0,.25) circle(1.5pt);
}}

\tikzset{
	MPSpairstaterotated1/.pic = {		
		\draw[thick] (.25,0.5) -- (0,.25)-- (0.25,0);
		\draw[thick,fill=myblue] (0.0,.25) circle(1.5pt);
}}
\tikzset{
	MPSpairstaterotated1solv/.pic = {		
		\draw[thick] (.25,0.5) -- (0,.25)-- (0.25,0);
}}

\tikzset{
	MPSpairstatefoldeddagrot/.pic = {		
		\draw[thick] (-.25,0.5) -- (0,.25)-- (-0.25,0);
		\draw[thick,fill=blue] (0.0,.25) circle(1.5pt);}}

\tikzset{
	MPSpairstatefoldeddagrotprime/.pic = {		
		\draw[thick] (-.25,0.5) -- (0,.25)-- (-0.25,0);
		\draw[thick,fill=orange] (0.0,.25) circle(1.5pt);}}

\tikzset{
	MPSpairstaterotated2/.pic = {		
		\draw[thick] (-.25,0.5) -- (0,.25)-- (-0.25,0);
		\draw[thick,fill=black] (0.0,.25) circle(1.5pt);
}}

\tikzset{
	connection/.pic = {		
		\draw[thick] (-.25,0.5) -- (0.0,.25)-- (0.25,.5);
}}
\newcommand*{\wcirc}{{\tikz[]{
			\draw[thick, fill=white] (0,0) circle (.1);}}\,}

\newcommand*{\bcirc}{{\tikz[]{
			\draw[thick, fill=black] (0,0) circle (.1);}}\,}

\newcommand*{\brec}{{\tikz[]{\draw[thick, fill=black](-0.1,-0.1)rectangle(.1,.1);}}\,}
\newcommand*{\arcstate}{{\tikz[baseline= 0]{\draw (0,0) arc
			[
			start angle=-90,
			end angle=90,
			x radius=.2,
			y radius =.1
			] ;
	}}\,}
\newcommand*{\wrec}{{\tikz[]{\draw[thick, fill=white](-0.1,-0.1)rectangle(.1,.1);}}\,}

\definecolor{lightgreen}{HTML}{90EE90}

\usetikzlibrary{shapes.geometric,calc,matrix,arrows,snakes,shapes,patterns}
\theoremstyle{plain}

\newtheorem{property}{Property}
\newtheorem{lemma}{Lemma}

\theoremstyle{definition}
\newcommand{\Op}[1]{\mathbb{#1}}

\newcommand{\dagg}{^\dagger}

\renewcommand{\d}{\text{d}}

\newcommand{\unity}{\mathds{1}}

\newcommand{\ra}{\rightarrow}

\theoremstyle{remark}

\newcommand{\unm}{\frac{1}{2}}

\usepackage{xcolor}
\hypersetup{
	colorlinks,
	linkcolor={blue!50!black},
	citecolor={blue!50!black},
	urlcolor={blue!80!black}
}

\let\svtikzpicture\tikzpicture
\def\tikzpicture{\noindent\svtikzpicture}

\usepackage{slashed}

\newcommand{\be}{ \begin{equation} }
	\newcommand{\ee}{\end{equation}}
\newcommand{\bea}{\begin{eqnarray}}
	\newcommand{\eea}{\end{eqnarray}}
\newcommand{\bse}{\begin{subequations}}
	\newcommand{\ese}{\end{subequations}}

\begin{document}	
	\title{Growth of entanglement of generic states under dual-unitary dynamics}
	
	\author{Alessandro Foligno}
	\affiliation{School of Physics and Astronomy, University of Nottingham, Nottingham, NG7 2RD, UK}
	\affiliation{Centre for the Mathematics and Theoretical Physics of Quantum Non-Equilibrium Systems,
		University of Nottingham, Nottingham, NG7 2RD, UK}
	\author{Bruno Bertini}
	\affiliation{School of Physics and Astronomy, University of Nottingham, Nottingham, NG7 2RD, UK}
	\affiliation{Centre for the Mathematics and Theoretical Physics of Quantum Non-Equilibrium Systems,
		University of Nottingham, Nottingham, NG7 2RD, UK}
	
	\begin{abstract}
		Dual-unitary circuits are a class of locally-interacting quantum many-body systems displaying unitary dynamics also when the roles of space and time are exchanged. These systems have recently emerged as a remarkable framework where certain features of many-body quantum chaos can be studied exactly. In particular, they admit a class of ``solvable" initial states for which, in the thermodynamic limit, one can access the full non-equilibrium dynamics. This reveals a surprising property: when a dual-unitary circuit is prepared in a solvable state the quantum entanglement between two complementary spatial regions grows at the \emph{maximal} speed allowed by the local structure of the evolution. Here we investigate the fate of this property when the system is prepared in a \emph{generic} pair-product state. We show that in this case the entanglement increment during a time step is sub-maximal for finite times, however, it approaches the maximal value in the infinite-time limit. This statement is proven rigorously for dual-unitary circuits generating high enough entanglement, while it is argued to hold for the entire class. 
	\end{abstract}	
	
	\maketitle
	
	
	\section{Introduction}	
	
	The evolution of quantum entanglement gives a universal and unifying characterisation of non-equilibrium dynamics in a wide range of quantum many-body systems ranging from lattice models to relativistic field theories~\cite{amico2008entanglement, calabrese2009introduction, calabrese2017lecture, laflorencie2016quantum}. Whereas the analysis of specific local observables is clouded by a plethora of system- and observable-specific effects, the evolution of entanglement over large scales does not depend on such inessential details and returns a clear portrait of the full (generalised) thermalisation process~\cite{calabrese2005evolution, nahum2017quantum}. Whenever a quantum many-body system with local interactions is prepared in an out-of-equilibrium state with low entanglement, and then let to follow its own unitary evolution, the entanglement between different spatial regions is observed to grow in time, signalling the proliferation of quantum correlations. In the course of this process the entanglement entropy of a given subsystem is transformed into \emph{thermodynamic entropy} and eventually saturates to a time-independent value indicating the onset of relaxation~\cite{deutsch2013microscopic, beugeling2015global, gurarie2013global, alba2017entanglement, calabrese2005evolution}. Unless specific competing mechanisms are introduced --- such as disorder~\cite{dechiara2006entanglement, znidaric2008many, nandkishore2015many}, confinement~\cite{kormos2017real}, or local measurements~\cite{skinner2019measurement, li2019measurement, vasseur2019entanglement} --- the entanglement grows linearly in time, irrespective of the nature of the system  dynamics~\cite{calabrese2005evolution, fagotti2008evolution, bertini2018entanglementand, alba2018entanglement, alba2017entanglement, alba2018entanglement,lagnese2021entanglement, laeuchli2008spreading, kim2013ballistic, pal2018entangling, bertini2019entanglement, piroli2020exact, gopalakrishnan2019unitary, nahum2017quantum, zhou2020entanglement, liu2014entanglement, asplund2015entanglement, vonkeyserlingk2018operator, klobas2021exact, klobas2021entanglement}. 
	
	The linear growth of entanglement naturally defines a velocity --- known as \emph{entanglement velocity}~\cite{kim2013ballistic, nahum2017quantum, vonkeyserlingk2018operator} --- which is obtained dividing the slope of the growth by the density of stationary entropy. The entanglement velocity is the key emergent parameter of the thermalisation process: it gives information on when subsystems start approaching stationarity and, at the same time, determines the feasibility of classical simulations of the quantum dynamics~\cite{schuch2008entropy, schuch2008on, perales2008entanglement, hauke2012can}. While it is clear that the entanglement velocity depends on geometry and couplings of a given system~\cite{alba2017entanglement, zhou2020entanglement, bertini2022growth}, it is less obvious whether it also depends on the initial configuration. One might expect that the dependence on the initial configuration should be mild, and all configurations leading to the same stationary state are characterised by the same entanglement velocity: some numerical observations supporting this expectation have been presented in Ref.~\cite{bertini2019entanglement}. On the other hand, the entanglement velocity describes a truly out-of-equilibrium regime taking place prior to relaxation and when a full scrambling of quantum information has yet to take place. For free systems for instance, initial configurations leading to the same stationary state can have different entanglement velocities~\cite{bertini2018entanglementand}. The same is expected for interacting integrable systems, where a formula for the entanglement velocity~\cite{alba2017entanglement, bertini2022growth} is only known for a special class of initial states~\cite{ghoshal1994boundary, piroli2017what}. These examples show that, at least for integrable models, the entanglement velocity contains more information than the stationary state and the intuitive expectation discussed above fails. For quantum chaotic systems, however, the question is still open.

	Here we analyse this question in the context of chaotic ``{local quantum circuits}", i.e., chains of qudits evolved by discrete applications of local unitary operators. These systems are useful idealisations of generic quantum matter and, over the last few years, have helped understanding information spreading~\cite{nahum2017quantum, nahum2018operator, vonkeyserlingk2018operator, chan2018solution, khemani2018operator, rakovszky2018diffusive, bertini2019entanglement, zhou2020entanglement}, spectral statistics~\cite{bertini2018exact, friedman2019spectral, bertini2018exact,chan2018spectral, flack2020statistics, bertini2021random, bertini2021random, fritzsch2021eigenstate, kos2021thermalization, bertini2022exact, garratt2021manybody, garratt2021local,chan2021many}, and thermalisation~\cite{piroli2020exact, klobas2021exact, klobas2021exactII, kos2021thermalization} in quantum many-body systems. Specifically, here we consider a particular class of local quantum circuits known as ``{dual unitary circuits}"~\cite{bertini2019exact}, which are defined by the property that their bulk dynamics remain unitary also when exchanging the roles of space and time. The most remarkable feature of these systems is that, despite being quantum chaotic, they allow for exact calculations of many relevant many-body quantities~\cite{gopalakrishnan2019unitary, bertini2020operator, bertini2020operator2, claeys2020maximum, reid2021entanglement, fritzsch2021eigenstate, claeys2021ergodic, suzuki2022computational, ho2022exact, ippoliti2022dynamical, kasim2022dual, claeys2022exact, jonay2021triunitary, milbdradt2022ternary}. Surprisingly, even the very quantum chaotic nature of dual-unitary circuits can be rigorously proven~\cite{bertini2018exact, bertini2021random}.  
	
	Dual-unitary circuits admit a class of ``{solvable}" initial states~\cite{bertini2019entanglement, piroli2020exact}, whose dynamics can be characterised exactly in the thermodynamic limit~\cite{bertini2019entanglement, piroli2020exact, claeys2022emergentquantum, ippoliti2022dynamical}. When evolving from solvable states dual-unitary circuits display \emph{maximal} entanglement growth, namely they show the largest entanglement growth compatible with the local structure of the time-evolution~\cite{bertini2019entanglement, piroli2020exact}. In fact, it has been recently shown in Ref.~\cite{zhou2022maximal} that such a maximal growth is only attainable in dual-unitary circuits. For generic initial states, however, dual-unitarity does not provide any obvious simplification and exact calculations fall out of reach. In addition, many of the special features of the dynamics of solvable states, including the maximal growth of entanglement, are observed to disappear in finite-time numerical experiments~\cite{bertini2019entanglement, piroli2020exact, ippoliti2022dynamical}. 
	
	Here we show that, remarkably, some exact statements can be made also for generic initial states. In particular, we consider the entanglement evolution from ``generic pair-product states", i.e., non-solvable states written as products of arbitrary two-site states, and show that the entanglement velocity is \emph{maximal} for \emph{almost all} dual-unitary circuits. Therefore, it is \emph{almost always} independent of the initial configuration.
	
	To find these results, we introduce spacetime-dependent noise that preserves dual-unitarity and show that the entanglement velocity averaged over the noise approaches the maximal value for large times. We then prove that this implies asymptotic maximality of the entanglement velocity for each realisation. Our statements are established rigorously for circuits made of dual-unitary gates with high enough ``entangling power", which measures how much a gate can entangle two qubits. These include dual-unitary gates constructed with complex Hadamard matrices~\cite{gutkin2020local} and four-leg perfect tensors~\cite{huber2018bounds, rather2022thirty}. We also we present a constructive way --- supported by numerical checks --- to extend them.  
	
	The rest of this paper is laid out as follows. In Sec.~\ref{sec:setting} we introduce the systems and initial states considered in this work. In Sec.~\ref{sec:entvel} we introduce the entanglement velocity, which is the quantity of interest, and review its calculation for dual-unitary circuits evolving from solvable states. Sec.~\ref{sec:randomDU} contains our main results: we begin by introducing the space-time dependent noise and show how maximality on average implies maximality for each single realisation. In Sec.~\ref{sec:bound} we bound from below the averaged entanglement entropy with a function depending on the gates solely through their entangling power. Then, in Sec.~\ref{sec:rigproof}, we prove maximality on average for circuits made of gates with large enough entangling power, while in Sec.~\ref{sec:genericgates} we argue that the proof can be extended to all dual-unitary circuits and in Sec.~\ref{sec:numericsSR} we present some supporting numerical evidence. Finally, in Sec.~\ref{sec:MPS} we show that our result is robust if one considers more general low-entangled initial states. Our conclusions and final remarks are reported in Sec.~\ref{sec:discussion}. The four appendices contain a number of complementary technical points.

	\section{Setting}
	\label{sec:setting}
	
	A one-dimensional local quantum circuit is a chain of $2L$ qudits --- with $d$ internal states ---  where the evolution occurs in discrete time-steps and describes local interactions. In particular, considering circuits where the time evolution is implemented in the so called ``brickwork" geometry, we write the unitary operator evolving the system from time $t$ to time $t+1$ as
	\be
	\Op{U}(t)=\Op{U}_{2}(t)\cdot\Op{U}_{1}(t)\,,
	\label{eq:U}
	\ee
	where we introduced 
	\begin{align}
		\label{eq:U1U2}
		\Op{U}_{1}(t)=\bigotimes_{x\in\mathbb Z_L} U_{x,t},\qquad \Op{U}_{2}(t)=\!\!\!\bigotimes_{x\in\mathbb Z_L+\unm}\!\!\! U_{x,t+1/2}.
	\end{align}
	The operator $U_{x,t}$ acts non-trivially, as the $d^2\times d^2$ unitary matrix $U({x,t})$, only on the qudits at positions $x$ and $x+1/2$. {The matrices $\{U(x,t)\}$ are known as ``local gates" and encode the physical properties of the system. In particular, whenever
		\be
		U(x,t) = U,\qquad \forall x,t,	
		\ee	
		the evolution operator is invariant under two-site shifts in time and space. We will refer to this case as a spacetime translational invariant quantum circuit.} 
	
	Note that in Eq.~\eqref{eq:U1U2} we labelled sites by half integers and assumed periodic boundary conditions so that the (half-odd) integers $x$ and $x+L$ denote the same site. We also remark that the form \eqref{eq:U} of the time-evolution operator implies that there is a strict maximal speed for the propagation of correlations. This means that any pair of local operators $a_x$ and $b_y$ evolved up to time $t$ satisfy
	\be
	[a_x(t),b_y(t)]=0, \qquad \abs{\ceil{x}-\ceil{y}}>2v_{\rm max}t,
	\ee
	where $\ceil{\bullet}$ denotes the ceiling function (smallest integer larger or equal to the argument). Moreover, our choice of units implies a maximal speed $v_{\rm max}=1$.

	We consider a particular class of local quantum circuits called dual-unitary circuits~\cite{bertini2019exact}. Their defining property is that they are generated by local gates that remain unitary under a particular reshuffling which corresponds to switching space and time. More precisely, defining a matrix $\tilde U$ with elements  
	\be
	{\tilde U}_{(j,l); (i,k)} = {U}_{(i,j);(k,l)}\,,\quad i,j,k,l=0,\ldots,d-1,
	\ee
	where we set $(i,j)=i*d+j$, we require 
	\be
	U^\dag  U= U U^\dag =I,\qquad\qquad \tilde U^\dag \tilde U=\tilde U\tilde U^\dag =I. 
	\label{eq:dualunitarity}
	\ee
	Whilst the first condition is the standard unitarity requirement for the local gate, the second one is imposing that the gate acts as a unitary matrix also when the roles of space and time are swapped. These constraints admit solution for all local Hilbert space dimensions $d\geq2$, however, a full parameterisation is only known for $d=2$~\cite{bertini2019exact, gutkin2020local, claeys2021ergodic, prosen2021many, bertini2021random, borsi2022remarks}. It is also useful to recall that, even though some of the solutions to \eqref{eq:dualunitarity} are integrable~\cite{bertini2020operator2, claeys2021ergodic, borsi2022remarks, giudice2022temporal}, i.e., generate evolution operators with an extensive number of local conserved charges, the integrable instances can only form a lower dimensional sub-manifold of the total manifold of dual-unitary circuits. This can be intuitively understood by noting that the two equations \eqref{eq:dualunitarity} are left invariant by the transformation 
	\be
	U\longmapsto   u_{+}\otimes u_{-} \cdot U \cdot v_{+}\otimes v_{-},
	\label{eq:DUpreservingtransformation}
	\ee
	with $u_{+},\ldots, v_{-}$ arbitrary elements of the group of $d\times d$ unitary matrices, which we denote by $U(d)$. This transformation is generally enough to break any non-trivial conservation law. In other words, dual unitary circuits are \emph{generally} non-integrable or quantum chaotic.
	
	{
		\subsection{Entangling Power}
		\label{sec:entpow}
		
		A feature of the local gate $U$ which will prove to be important in the following is its \emph{entangling power}. The latter is a measure of the average entanglement produced by $U$ when acting on Haar-random product states, see, e.g., Ref.~\cite{rather2020creating}. In particular, as shown in Refs.~\cite{rather2020creating, aravinda2021from, rather2022construction}, for dual-unitary circuits it can be expressed as 
		\be
		p = \frac{d^4-{\rm tr}[({\tilde U}^{t_2} ({\tilde U}^{t_2})^\dag)^2]}{d^2(d^2-1)},
		\label{eq:defp}
		\ee
		where $(\cdot)^{t_2}$ denotes the partial transpose with respect to the second qudit. From \eqref{eq:defp} one can immediately verify that $p$ is invariant under \eqref{eq:DUpreservingtransformation}.  
		
		As we recall in Appendix~\ref{app:avegate}, the entangling power~\eqref{eq:defp} fulfils     
		\be
		0\leq p \leq 1.
		\ee 
		The lower bound is attained when ${\tilde U}^{t_2} ({\tilde U}^{t_2})^\dag/d^2$ is a rank-1 projector. This happens when, up to the transformation \eqref{eq:DUpreservingtransformation}, $U$ coincides with the SWAP gate. Namely, it merely swaps the states of the two qudits it acts on, generating no entanglement. Instead, the upper bound is attained for
		\be
		{\tilde U}^{t_2} ({\tilde U}^{t_2})^\dag =  ({\tilde U}^{t_2})^\dag {\tilde U}^{t_2} = I.
		\label{eq:Urequirement}
		\ee
		To understand this condition it is useful to think of $U$ a state of four qudits with amplitudes $\{U_{(a_1,a_2); (a_3,a_4)}\}$. In this language, Eq.~\eqref{eq:Urequirement} means that the subset formed by the first and fourth qudits is maximally entangled with its complement. Recalling that the gate $U$ also fulfils \eqref{eq:dualunitarity} we see that also the subsets formed by first and second and first and third qubits are maximally entangled with their complements. In fact, in the state defined by $U$ any bipartition of the four qudits has maximal entanglement with its complement. Gates generating states with this property are called \emph{perfect tensors}~\cite{Pastawski:2015qua} or 2-unitary gates~\cite{Karol2015}. Perfect tensors with four entries exist for every local Hilbert space dimension strictly larger than $d=2$~\cite{huber2018bounds, rather2022thirty} and, therefore, for $d>2$ the upper bound $p=1$ can be attained. Instead, for $d=2$ the maximal value that $p$ can attain is~\cite{bertini2019exact}   
		\be
		p=\frac{d}{d+1}\,,
		\ee
		and, up to \eqref{eq:DUpreservingtransformation}, it is attained by local gates of the form
		\be
		U_{(i,j),(k,l)}=\delta_{il}\delta_{jk}\exp({\rm i}\frac{2\pi ij}{d}).
		\label{eq:Hadamardgates}
		\ee
		The family \eqref{eq:Hadamardgates} of dual-unitary gates has been constructed in Ref.~\cite{gutkin2020local} using complex Hadamard matrices. Here, for brevity, we call it the ``Hadamard family''. 
	}

	\subsection{Quantum Quench}	
	To study the out-of-equilibrium dynamics of the circuits we consider a standard quantum quench protocol: we prepare them in a non-equilibrium state $\ket{\Psi_0}$ and let them evolve according to their time-evolution operator. In particular, we consider generic ``pair-product" states of the form
	\be
	\ket{\Psi_0}=\frac{1}{d^{L/2}}\bigotimes_{x=1}^L \left(\sum_{i,j=0}^{d-1} m_{i;j} \ket{i,j}    \right),
	\label{eq:initialstate}
	\ee
	where $\{\ket{i}\}_{i=0}^d$ is a basis of the local Hilbert space and matrix $m$, with elements $m_{i;j}$, fulfils 
	\be
	\tr(m m\dagg)=d,
	\label{eq:normalization}
	\ee
	which ensures that $\ket{\Psi_0}$ is normalised to one. Apart from this condition, the matrix $m$ is \emph{completely generic}. 
	
	Although the family \eqref{eq:initialstate} does not represent the most general low-entangled state, it contains many physically relevant points --- in particular it contains all translational invariant product states --- and it is complex enough to show generic behaviour. In most of the paper we focus on this family to reduce to a minimum the technical complications, while in Sec.~\ref{sec:MPS} we show that our techniques can be applied to more general families of matrix product states leading to qualitatively similar results.
	
	The evolution quantum circuits can be conveniently represented graphically. One depicts the local gates as a four leg tensors
	\begin{align}
		{U}_{(i,j);(k,l)} =
		\begin{tikzpicture}[baseline=(current  bounding  box.center),scale=1.5]	
			\path (0,0) pic[scale=2]{U};
			\node at (-0.35,-0.15) {$i$};
			\node at (0.35,-0.15) {$j$};
			\node at (-0.35,.8) {$k$};
			\node at (0.35,.8) {$l$};
		\end{tikzpicture},
		\label{eq:Upic}
	\end{align}
	and the initial state matrix $m$ as a two-leg one
	\be
	m_{i;j} =\scalebox{1}{\begin{tikzpicture}[baseline=(current  bounding  box.center),scale=1]	
			\foreach \i in {0}
			{		\path (\i ,0) pic[scale=2	]{MPSpairstate};}
			\node at (0.65,1.1) {$j$};
			\node at (-0.6,1.1) {$i$};
			\,.
	\end{tikzpicture}}.
	\label{eq:mpic}
	\ee
	When it does not lead to confusion the indices can be dropped to represent the actual tensor instead of its elements. For instance, \eqref{eq:dualunitarity} are conveniently represented as 
	\begin{align}
		\begin{tikzpicture}[baseline=(current  bounding  box.center),scale=1.5]	
			\draw[thick] (0,0) pic[scale=1.5]{Udag};
			\path (0,.6) pic[scale=1.5]{U};
			\draw[thick] (-.25,.495)--(-.25,.605);
			\draw[thick] (.25,.495)--(.25,.605);
		\end{tikzpicture} &= \begin{tikzpicture}[baseline=(current  bounding  box.center),scale=1.5]	
			\draw[thick] (0,0) pic[scale=1.5]{U};
			\path (0,.6) pic[scale=1.5]{Udag};
			\draw[thick] (-.25,.495)--(-.25,.605);
			\draw[thick] (.25,.495)--(.25,.605);
		\end{tikzpicture}=\begin{tikzpicture}[baseline=(current  bounding  box.center),scale=1.5]	
			\draw[thick] (-.25,0)--(-.25,1.1);
			\draw[thick] (.25,0)--(.25,1.1);
		\end{tikzpicture} \,,\label{eq:unitarity}\\
		\notag\\
		\begin{tikzpicture}[baseline=(current  bounding  box.center), scale=1.5]	
			\draw[thick] (0,.5) pic[scale=1.5, rotate=180]{Udag};
			\path (.6,0) pic[scale=1.5]{U};\draw[thick] (.245,0)--(.355,0);
			\draw[thick] (.245,0.5)--(.355,0.5);
		\end{tikzpicture} & = \begin{tikzpicture}[baseline=(current  bounding  box.center), scale=1.5]	
			\draw[thick] (0,0) pic[scale=1.5]{U};
			\path (.6,0.5) pic[scale=1.5,rotate=180]{Udag};\draw[thick] (.245,0)--(.355,0);
			\draw[thick] (.245,0.5)--(.355,0.5);
		\end{tikzpicture} = \begin{tikzpicture}[baseline=(current  bounding  box.center),scale=1.5]	
			\draw[thick] (-.05,0.225) -- (.805,.225);
			\draw[thick] (-.05,-0.225) -- (.805,-.225);
		\end{tikzpicture}
		\label{diag:dualunitarity}\,,
	\end{align}
	where we introduced the diagram
	\be
	{U}^\dag =
	\begin{tikzpicture}[baseline=(current  bounding  box.center),scale=1.5]	
		\path (0,0) pic[scale=2]{Udag};
		\node at (0,-.05) {};
	\end{tikzpicture}.
	\ee
	Instead, the state at time $t$ is depicted as
	\be
	\Op{U}^t \ket{\Psi_0}= \scalebox{1.1}{\begin{tikzpicture}[baseline=(current  bounding  box.center),scale=1]
			\foreach \j in {0,...,1}
			{			\draw[thick] (1.75+0.070710678118
				,\j+0.7-0.029289321881) arc (90:225:0.1);
				\draw[thick] (1.75,\j) arc (135:270:0.1);
				\draw[thick] (5.75,\j) arc (135:-90:0.1);
				\draw[thick] (5.75,\j+0.5) arc (225:450:0.1);
				\draw[dashed, opacity=.8] (1.8,0.67+\j)--(5.8,0.67+\j);
				\draw[dashed, opacity=.8] (1.8,0.33-.5+\j)--(5.8,0.33-.5+\j);
			}		
			\foreach \i in {3,...,6}
			{	 \path (-0.5+\i,-0.5) pic{MPSpairstate};}
			\foreach \i in {2,...,5}
			{\foreach \j [evaluate=\j as \jn using {mod(\j,2)}]  in {0,...,3} 
				{\draw(\i+\jn/2,\j/2) pic{U};
				}
			}           
	\end{tikzpicture}}\,, 
	\label{eq:stateevorep}
	\ee
	where we took $L=4$ and $t=2$, we represented the matrix multiplication from bottom to top and we conveniently depicted the space-time translational invariant case  with periodic boundary conditions. The representation in Eq.~\eqref{eq:stateevorep} makes it clear why this particular way of applying local gates is called brickwork geometry. 	
	
	\section{Entanglement Growth}
	\label{sec:entvel}	
	
	In this paper we are interested in the evolution of the entanglement between a contiguous block of $2L_A$ qudits, $A=\{x_1, x_1+1/2,\ldots x_2\}$, and its complement, $\bar A = \{1/2,\ldots,L\}\setminus A$, in the state \eqref{eq:stateevorep}. In particular, we will focus on the regime where the entanglement typically grows linearly in time~\cite{calabrese2005evolution, fagotti2008evolution, bertini2018entanglementand, alba2018entanglement, alba2017entanglement, alba2018entanglement,lagnese2021entanglement, laeuchli2008spreading, kim2013ballistic, pal2018entangling, bertini2019entanglement, piroli2020exact, gopalakrishnan2019unitary, nahum2017quantum, zhou2020entanglement, liu2014entanglement, asplund2015entanglement, vonkeyserlingk2018operator, klobas2021exact, klobas2021entanglement}, i.e., 
	\be
	2 v_{\rm max} t \leq  L_A \leq  L - L_A\,. 
	\label{eq:regime}
	\ee
	Since we are considering systems with local interactions, the entanglement between $A$ and $\bar A$ is produced starting from the boundaries between the two sub-systems. In the regime of interest the two boundaries between $A$ and $\bar A$ are causally disconnected and give identical contributions.
	
	To quantify the entanglement of the bipartition we compute the reduced density matrix
	\be
	\rho_A(t)=\tr_{\bar A}[\Op{U}^t \ketbra{\Psi_0}{\Psi_0} \Op{U}^{-t}]\,,
	\ee
	and evaluate its R\'enyi entropies 
	\be
	S_A^{(\alpha)}\!(t)=\frac{1}{1-\alpha} \log\tr[\rho_A(t)^\alpha],\qquad \alpha\in\mathbb R.
	\label{eq:renyi}
	\ee
	Note that $S_A^{(\alpha)}\!(t)$ is non-increasing in $\alpha$
	\be
	S_A^{(\alpha)}\!(t) \leq S_A^{(\beta)}\!(t), \qquad \alpha\geq \beta\,,  
	\label{eq:nondec}
	\ee
	and its liming value for $\alpha\to1$ corresponds to the celebrated \emph{entanglement entropy}
	\be
	\!\!\lim_{\alpha\to1}S_A^{(\alpha)}\!(t) \!=\! - \tr[\rho_A(t)\! \log\rho_A(t)]\!\equiv\! S_A(t),
	\ee
	which is a bona fide measure of bipartite entanglement~\cite{amico2008entanglement}.

	To analyse R\'enyi entropies in the regime \eqref{eq:regime} we note that the reduced density matrix can be represented as
	\begin{widetext}
		\begin{align}
			\rho_A(t)=\! 
			\frac{1}{d^{2t+L_A}}\scalebox{.7}{\begin{tikzpicture}[baseline=(current  bounding  box.center)]
					\foreach \i in {0,...,13}
					{	 \path (-0.5+\i,-0.5) pic{MPSpairstate};
						\path (-0.5+\i,-1+8) pic{MPSpairstateupsidedown};
					}
					\foreach \i in {0,...,5}
					{
						\draw (-.75+\i/2,+\i/2)--(-.75+\i/2,7-\i/2);		
						\draw (12+.75-\i/2,+\i/2)--(12+.75-\i/2,7-\i/2);
					}
					\foreach \i in {0,...,5}
					{\foreach \j in {0,...,\i}
						{\draw(\i-\j/2,\j/2) pic{U};
							\draw(\i-\j/2,6.5-\j/2) pic{Udag};
						}
					}        
					\foreach \i in {0,...,6}
					{  \foreach \j in {0,...,5}
						{			\draw(3.5+\i+\j/2,2.5-\j/2) pic{U};
							\draw(3.5+\i+\j/2,4+\j/2) pic{Udag};
						}		
					}   	
			\end{tikzpicture}},
			\label{eq:reducedAdiag}
		\end{align}
	\end{widetext}
	where we used the two-site product form of the initial state \eqref{eq:initialstate}, the normalisation \eqref{eq:normalization}, the unitarity of the gates $U$, and we introduced the diagram
	\be
	m^\dag =\scalebox{1}{\begin{tikzpicture}[baseline=(current  bounding  box.center),scale=1]	
			\foreach \i in {0}
			{		\path (\i ,0) pic[scale=2]{MPSpairstateupsidedown};}
	\end{tikzpicture}}.
	\ee
	Using the representation \eqref{eq:reducedAdiag} and employing the unitarity relations \eqref{eq:unitarity} one can readily show that if 
	\be
	2t<|\ceil*{x_2}-\ceil*{x_1}|,
	\ee
	the traces of powers of the reduced density matrix $\rho_A(t)$ factorise as follows
	\begin{align}
		\!\!\!\!\Tr[\rho_A(t)^\alpha]=\Tr[(C_{2t_{x_2}}\dagg C^{\phantom{\dag}}_{2t_{x_2}})^\alpha] \Tr[(C_{2t_{x_1}}\dagg C^{\phantom{\dag}}_{2t_{x_1}})^\alpha]. \label{eq:red_den_mat_decomposition}
	\end{align}
	Here $t_{x_i}=t-\{x_i\}$ ($\{\bullet\}\equiv \ceil{\bullet}-\bullet$ is the fractional part) and $C_x$ is a $d^{x}\times d^{x}$ matrix corresponding to the following diagram
	\be
	[C_{x}]_{a;b} =\frac{1}{d^{\frac{x}{2}}} \begin{tikzpicture}[baseline=(current  bounding  box.center),scale=1]	
		\foreach \i in {0,...,4}
		{\path (\i ,0) pic{MPSpairstate};
		}
		\foreach \i in {0,...,3}
		{\foreach \j in {0,...,\i}
			{\path (2-\i/2+\j ,2-\i/2) pic{U};
			}
		}		
		\node[scale=1] 	at (0+2.5,5.3-2.5) {$b_x$};
		\node[scale=1] 	at (1+2.5,4.3-2.5) {$\ddots$};
		\node[scale=1] 	at (2+2.5,3.3-2.5) {$b_1$};
		\node[scale=1] 	at (2-.4,5.3-2.6) {$a_x$};
		\node[scale=1, rotate=90] 	at (1-.4,4.3-2.6) {$\ddots$};
		\node[scale=1] 	at (-.4,3.3-2.6) {$a_1$};
	\end{tikzpicture},
	\label{diag:Ct}
	\ee
	where $q_j$ denotes the $j$-th digit of $q$ in base $d$. Thanks to the unitarity of the local gates, we have the condition
	\be
	{\rm Tr}[ C^{\phantom{\dag}}_x C_x^\dag]=1, \qquad \forall x\,.
	\label{eq:traceCCdag}
	\ee
	To simplify the notation, from now on we assume $x_1,x_2$ to be integers. Plugging \eqref{eq:red_den_mat_decomposition} into \eqref{eq:renyi}, we can express the R\'enyi entropies as
	\be
	\label{eq:RenyiCt}
	S_A^{(\alpha)}\!(t)=\frac{2}{1-\alpha}\log {\rm Tr}[\left(C^{{\dag}}_{2t} C_{2t}^{\phantom{\dag}} \right)^\alpha],
	\ee
	where the factor of $2$ occurs because the two independent boundaries between $A$ and $\bar A$ give the same contribution. 
	
	Since $C^{\phantom{\dag}}_x C_x^\dag $ is Hermitian, positive definite, and fulfils \eqref{eq:traceCCdag} it is easy to find a bound for the powers of its trace. To see this, we diagonalise the matrix and express the above conditions in terms of its eigenvalues $\lambda_i$ as follows 
	\begin{align}
		\lambda_i\ge 0,\qquad\qquad \sum_{i=1}^{\mathcal N} \lambda_i=1,
		\label{eq:eq1}
	\end{align}
	where $\mathcal N=d^x$ is the dimension of the vector space on which the matrix acts. The constraints \eqref{eq:eq1} on generic real numbers lead to the following bound
	\begin{align}
		\frac{1}{{\mathcal N}^{\alpha-1}}\leq \Tr[(C_x\dagg C_x^{\phantom{\dag}})^{\alpha}]=\sum_i \lambda_i^{\alpha}\leq 1.\qquad \forall \alpha \ge1
		\label{eq:boundpowertrace}
	\end{align}
	Using this in \eqref{eq:RenyiCt}, we find 
	\be
	0 \leq  S_A^{(\alpha)}\!(t)\leq 4 t \log d\,,\qquad \forall\alpha \ge 1.
	\label{eq:Sbound}
	\ee
	The lower bound is reached when $C_x^\dag C^{\phantom{\dag}}_x$ is a projector on a one-dimensional space, while the upper bound is attained when it is maximally mixed, i.e.  
	\be
	C^{\phantom{\dag}}_x C_x^\dag = \frac{\unity_{x}}{d^{x}},
	\label{eq:upbound}
	\ee
	where $\unity_x$ is the identity matrix on $x$ qudits.

	We are now in a position to introduce the quantity of interest in this paper, i.e., the \emph{entanglement velocity}, which quantifies the asymptotic growth of entanglement in the out-of-equilibrium regime \eqref{eq:regime}. In our setting this quantity is defined as the ratio between half of the asymptotic slope of the entanglement entropy and the density of entropy of the stationary state --- the additional factor of two is included to isolate the the entanglement growth from a single boundary between $A$ and $\bar A$. More formally, for a circuit without local conservation laws, we have 
	\be
	v_{\rm E}\equiv \limsup_{t\to\infty}  \lim_{L_A\to\infty} \lim_{L\to\infty} \frac{S_A(t)}{4 t\log d}.
	\label{eq:velocityent}
	\ee
	Note that in \eqref{eq:velocityent} we used that the circuit has no local conservation laws to find the  density of its thermodynamic entropy ($2 \log d$) and we introduced the limit superior rather than the regular limit to make sure that $v_{\rm E}$ always exists. Analogously, we can introduce entanglement velocities for all R\'enyi entropies
	\be
	v_{\rm E,\alpha}\equiv \limsup_{t\to\infty}  \lim_{L_A\to\infty} \lim_{L\to\infty} \frac{S^{(\alpha)}_A(t)}{4 t\log d}, \qquad v_{\rm E,1}=v_{\rm E}\,.
	\label{eq:velocityentalpha}
	\ee
	Using \eqref{eq:nondec} and \eqref{eq:Sbound} we find the following general bound 
	\be
	0\leq v_{\rm E,\beta} \leq v_{\rm E,\alpha} \leq 1, \qquad \beta \geq \alpha. 
	\label{eq:boundvelocity}
	\ee
	Note that, up to now, we did not use the dual-unitarity of the gates at any point in the reasoning and, in fact, our discussion applies to any chaotic local quantum circuit. This is because generic matrices $m$ ``break" the special dual-unitarity property of the local gates, preventing any direct simplification. On the other hand, as we shall now see, for a special class of compatible matrices,  dual-unitarity immediately leads to an explicit expression for $S_A^{(\alpha)}\!(t)$.

	Let us consider a sub-class of pair-product states~\eqref{eq:initialstate} characterised by \emph{unitary} matrices $m$, i.e., matrices fulfilling the diagrammatic relations 
	       \be
		\begin{tikzpicture}[baseline=(current  bounding  box.center),scale=1.5]	
			\draw[thick, rounded corners] (-.25,.5) -- (.,.25) -- (.25,.5) -- (.5,.25) -- (.75,.5);
			\draw[thick,fill=myblue] (0,.25) circle	(.07);
			\draw[thick,fill=red] (0.5,.25) circle	(.07);
		\end{tikzpicture}
		=
		\begin{tikzpicture}[baseline=(current  bounding  box.center),scale=1.5]	
			\draw[thick, rounded corners] (-.25,.5) -- (.,.25) -- (.25,.5) -- (.5,.25) -- (.75,.5);
			\draw[thick,fill=red] (0,.25) circle	(.07);
			\draw[thick,fill=myblue] (0.5,.25) circle	(.07);
		\end{tikzpicture}= \begin{tikzpicture}[baseline=(current  bounding  box.center),scale=1.5]	
			\draw[thick, rounded corners] (-.25,.5) -- (.75,.5);
		\end{tikzpicture}\,.
		\label{eq:unitarym}
		\ee
	Repeatedly applying \eqref{eq:unitarym} and \eqref{diag:dualunitarity} we can fully contract the tensor network
	\be 
	C\dagg_x C^{\phantom{\dag}}_x = \frac{1}{d^x}
	\begin{tikzpicture}[baseline=(current  bounding  box.center),scale=1]	
		\foreach \i in {0,...,4}
		{\path (\i ,0) pic{MPSpairstate};
			\path (-0.5,\i+.5) pic{MPSpairstaterotated1};
			
		}
		\foreach \i in {0,...,3}
		{\foreach \j in {0,...,\i}
			{\path (2-\i/2+\j ,2-\i/2) pic{U};
				\path (1.5-\i/2 ,2.5-\i/2+\j) pic{Udag2};
			}
		}		
	\end{tikzpicture},
	\ee
	and find 
	\be
	C\dagg_x 	C^{\phantom{\dag}}_x=\frac{\unity_x}{d^x}.
	\label{diag:Ctsolvable}
	\ee
	Pair-product initial states with this property are part of a larger family of exactly treatable states, generically in MPS form, called \emph{solvable states}~\cite{piroli2020exact}.

	We see that, for solvable pair product states, $C\dagg_x C^{\phantom{\dag}}_x$ takes the maximally mixed form \eqref{eq:upbound}, therefore the entropies saturate the bound~\eqref{eq:Sbound}, i.e.,   
	\be
	S_A^{(\alpha)}\!(t)= 4 t \log d\,,\qquad \forall\alpha,
	\ee
	or, equivalently, all entropies have the \emph{maximal} increment over a time step
	\be
	\Delta S_A^{(\alpha)}(t) \equiv \frac{S_A^{(\alpha)}\!(t)-S_A^{(\alpha)}\!(t-1)}{4\log d} = 1,\qquad \forall\alpha.
	\label{eq:entanglementincrement}
	\ee
	This condition characterises all solvable states for large enough subsystems~\cite{piroli2020exact}. A particular consequence of this is also 
	\be
	v_{\rm E}=1. 
	\ee
	The goal of this paper is to show that, even if hidden, an effect of dual-unitarity is also present for generic initial states. As a consequence, even if \eqref{eq:entanglementincrement} does not hold at finite times, the entanglement velocity \emph{remains maximal}.

	\section{Entanglement Velocity from Generic Pair-Product States}
	\label{sec:randomDU}
	
	Our strategy to treat generic initial states is to introduce spacetime-dependent noise and  to show that our statements hold for arbitrary distributions of the noise. {More specifically, we consider a spacetime translational invariant, dual-unitary circuit characterised by a local gate $U$, and insert uncorrelated Haar-distributed ${U}(d)$ noise at each spacetime point through the transformation~\eqref{eq:DUpreservingtransformation}}. Note that the family of random dual-unitary gates produced in this way is the direct generalisation for generic $d$ of the family introduced in in the case $d=2$~\cite{bertini2020scrambling}.
	
	In this random setting it is natural to consider the averaged R\'enyi entropies 
	\be
	\bar S_A^{(\alpha)}(t) = \frac{1}{1-\alpha}\mathbb{E}[\log\tr(\rho_{A}(t)^\alpha)],
	\label{eq:averagedrenyi}
	\ee
	where $\mathbb{E}[\cdot]$ is the average over the set of unitaries 
	\be
	\label{eq:uvmats}
	\boldsymbol u\equiv \{u(\tau,x)\}_{\tau={1}/{2},1,\ldots, t; x={1}/{2},1,\ldots,L}\in U^{4Lt}(d),
	\ee
	and $U^x(d)$ denotes the direct product of $x$ copies of $U(d)$. 
	
	Analogously, we define the averaged entanglement velocities as 
	\be
	\bar v_{\rm E,\alpha}\equiv \limsup_{t\to\infty}  \lim_{L_A\to\infty} \lim_{L\to\infty} \frac{\bar S_A^{(\alpha)}\!(t)}{4 t\log d}, \qquad \bar v_{\rm E}=\bar v_{\rm E,1}\,.
	\label{eq:averagedvelocities}
	\ee
	With these definitions at hand, we are now in a position to state our main objective. Our goal is to prove the following property.
	\begin{property}
		\label{prop:P1}
		For 
		\emph{all} states of the form \eqref{eq:initialstate} 
		\be
		\bar v_{\rm E}=1\,.
		\label{eq:maxaveragedspeed}
		\ee
	\end{property}
	
	Before approaching the proof, let us analyse its implications. Recalling the bound in Eq.~\eqref{eq:boundvelocity} we see that this property implies that the average entanglement velocity is maximal for any initial state~\eqref{eq:initialstate}. Since we find our bound saturated on average, we intuitively expect the entanglement velocity to be maximal for almost every choice of the unitaries $\boldsymbol u$, i.e.\ in the non-random case. To make this statement more precise, consider the function $f$ whose limit superior for $t\to\infty$ is the entanglement velocity
	\begin{align}
		f(t,\{{u}_\pm({t,x})\},\{{v}_\pm({t,x})\})\equiv \lim_{L_A\to\infty} \lim_{L\to\infty} \frac{S_A(t)}{4t \log(d)}.
	\end{align}
	Note that, for any choice of the gates we have 
	\begin{align}
		f(t,\{{u}_\pm({t,x})\},\{{v}_\pm({t,x})\})\in[0,1].
		\label{eq:fbound}
	\end{align}
	Here we make this function depend on an semi-infinite square grid of gates, labelled by $(x,t)$, with $t=1,2,\ldots\infty$ and $x=-\infty,\ldots\infty$ (at finite times $t$ it actually depends only on a finite subset of such gates). This function is measurable for any $t$ because it is continuous \cite{tao2013introduction}: its associated measure $\Omega$ is the product of the Haar measures, { we call them $\Omega_{\rm loc}$}, of each unitary ${u}(t,x)$. Importantly, the measure $\Omega$ is fixed and does not depend on $t$ because it is a countable product of Haar measures on the semi-infite square grid described above. Since $f$ is positive and measurable for any $t$, we can apply Fatou's lemma~\cite{tao2013introduction} and exchange the order between limsup and integral
	\begin{align}
		\!\!\int \!\!\d \Omega\,  v_{\rm E} = \!\!\int \!\!\!\d \Omega\, \:\limsup_{t\ra\infty}	f
		\ge \limsup_{t\ra\infty} \int \!\!\!\d \Omega \:	f = \bar v_{\rm E} =1,
	\end{align}
	where, in the last equality, we used Property \ref{prop:P1}. Using the bound \eqref{eq:fbound}, we then find
	\begin{align}
		\int \!\!\d \Omega\,  v_{\rm E} =1.
		\label{eq:avevmax}
	\end{align}
	Since this saturates the bound on the velocity, it is implied that, for almost all choices of gates, the entanglement velocity is $1$, i.e., 
	\begin{align}
		\bar v_{\rm E}=1\quad\implies\quad v_{\rm E}\approx 1.
	\end{align}
	Here the symbol $\approx$ indicates that the equality holds for \emph{almost all} choices of gates.

	{ Importantly, the statement \eqref{eq:avevmax} has strong implications also for the non-random case. Indeed, for any given non-random distribution of the one-site gates $\boldsymbol u$ --- for instance one that is uniform in space and time --- one can consider adding an arbitrary small distribution of ``noise". Namely, one modifies the one-site gates as  
		\begin{align}
			u^{(\epsilon)}(\tau,x)
			:=u(\tau,x)\cdot w^{(\epsilon)}(\tau,x),
			\label{eq:epsilonball}
		\end{align}
		where, for each $(\tau,x)$, $w^{\epsilon}(\tau,x)$ is a unitary matrix extracted at random from 
		an $\epsilon-$ball centred on the identity matrix. We emphasise that this ball is taken to have unit measure. For instance, one can take $w^{\epsilon}(\tau,x)$ distributed according to a Gaussian measure  
		\be
{\rm d}\Omega_{\rm loc} \mapsto {\rm d}\Omega_{\rm loc}^{(\epsilon)}= 
\prod_{a=1}^{d^2-1}\frac{1}{\sqrt{2\pi}\epsilon}
\exp\left(-{\frac{1}{2}} \frac{\theta_a^2}{\epsilon^2}\right){\rm d}\boldsymbol \theta\,,
\ee
or a box-measure
\be
{\rm d}\Omega_{\rm loc}\mapsto {\rm d}\Omega_{\rm loc}^{(\epsilon)}= 
\prod_{a=1}^{d^2-1} \frac{1}{2\epsilon}
\Theta(\epsilon-|\theta_a|){\rm d}\boldsymbol \theta\,,
\ee
where ${\boldsymbol \theta=\{\theta_a\}_{a=1}^{d^2-1}}$ are the Euler angles specifying $w^{\epsilon}(\tau,x)$. 

The choice \eqref{eq:epsilonball} implies that, by choosing small enough $\epsilon$, one can make $u^{(\epsilon)}(\tau,x)$ arbitrarily close to $u(\tau,x)$. Then, Eq.~\eqref{eq:avevmax} guarantees that for every $\epsilon>0$ the entanglement velocity averaged over $u^{(\epsilon)}(\tau,x)$ is maximal, i.e., \emph{it is maximal when we get arbitrarily close to the non-random case}.}
	
	In the upcoming subsections we prove Property~\ref{prop:P1}. In Sec.~\ref{sec:bound} we show that $\bar S_A(t)$ can be bounded from below in terms of a function depending on the gates only through the entangling power of $U$. In Sec~\ref{sec:rigproof} we show that for   
	\be
	p \geq \bar p(d) \equiv \frac{d^2-1}{d^2} \left(1-\frac{1}{\sqrt{2d+2}}\right)\,,
	\label{eq:pbound}
	\ee
	this bound leads to a rigorous proof of Property~\ref{prop:P1}. In particular, recalling Sec.~\ref{sec:entpow} and noting that 
	\be
	\bar p(d) < \frac{d}{d+1}< 1,\label{eq:entanglingpowbound}
	\ee
	we prove that Property~\ref{prop:P1} holds for perfect tensors and for the Hadamard family (cf. Eq.~\eqref{eq:Hadamardgates}) while in Subsection~\ref{sec:genericgates} we argue that Property~\ref{prop:P1} can be extended to all $p$ except for a neighbourhood of $p=0$. Instead, in Sec.~\ref{sec:numericsSR} we present numerical evidence supporting the claim that $v_{\rm E}$ is one for concrete individual realisations of the noise. Finally,  in Sec.~\ref{sec:MPS} we show that  similar statements can be made for a more general class of initial states in MPS form.

	\subsection{Bound on $\bar S_A(t)$}
	\label{sec:bound}
	
	We aim to bound $\bar S_A(t)$ in the regime \eqref{eq:regime} by a function depending on the local gates only through $p$. We begin by using \eqref{eq:nondec} and \eqref{eq:Sbound} which give  
	\be
	\bar S_A^{(2)}(t) \leq \bar S_A(t) \leq 4 t \log d,	 
	\ee
	Noting now that the function
	\be
	f(x)= -{2}\log x,
	\ee
	is convex, we have 
	\be
	-{2\log \mathcal P_{2t}} \leq \bar S_A^{(2)}(t),	
	\label{eq:boundS2bar2}
	\ee
	where we introduced the averaged purity for the matrix $ C_x^\dag C_x^{\phantom{\dag}}$
	\be
	\mathcal P_x \equiv  \mathbb{E}[{\rm Tr}[(C_x^\dag C^{\phantom{\dag}}_x)^2]]. 
	\ee
	Putting all together we have 
	\be
	-{2\log \mathcal P_{2t}} \leq \bar S_A(t) \leq 4 t \log d.	 
	\label{eq:boundS2bar}
	\ee	
	{

		To conclude, we show that $\mathcal P_x$ depends on the local gates only through their entangling power. To this end, we note that, since $\boldsymbol u$ are independently distributed at each spacetime point, the average $\mathbb{E}[\cdot]$ factorises on each separate gate. This allows us to adopt a convenient tensor-network representation for $\mathcal P_x$, which is obtained by folding the four diagrams for $C_x, C\dagg_x,C_x$, and $C\dagg_x $ on top of each other and averaging (see Ref \cite{bertini2020scrambling} for a more detailed explanation of this ``folded" diagrammatic representation)
		\be
		\!\!\mathcal P_x = \begin{tikzpicture}[baseline=(current  bounding  box.center),scale=1]	
			\foreach \i in {0,...,4}
			\path (\i ,0) pic{MPSpairstatef};
			\foreach \i in {0,...,3}
			{\foreach \j in {0,...,\i}
				{\path (2-\i/2+\j ,2-\i/2) pic{Uffolded};
					\draw[thick, fill=white] (2-\i/2-0.25 ,2-\i/2+0.5) pic{circ2};
				}
				\draw[thick,fill=white] (2-\i/2+\i+0.25,2-\i/2+0.5) pic{rec2};
			}
			\foreach \i in {0,...,4}
			{
				\draw[thick,fill=white] (2-\i/2+\i+0.25,2-\i/2+0.5) pic{rec2};
				\draw (2-\i/2-0.25 ,2-\i/2+0.5) pic{circ2};
			}
			\draw [decorate, decoration = {brace}]   (2.5,2.75) -- (4.75,0.5);
			\node[scale=1] 	at (4,1.7) {$x$};
		\end{tikzpicture},
		\label{eq:Pt}
		\ee
		where we introduced the vectors 
		\be
		\begin{aligned}
			\ket{\wcirc}&\equiv \frac{1}{d}\sum_{i,j,k,l} \delta_{ij}\delta_{kl}\ket{ijkl}\,,\\
			\ket{\wrec}&\equiv \frac{1}{d}\sum_{i,j,k,l} \delta_{il}\delta_{jk}\ket{ijkl}\,,
			\label{eq:circrec}
		\end{aligned}
		\ee
		and the averaged gate
		\begin{align}
			&W=\begin{tikzpicture}[baseline=(current  bounding  box.center),scale=1.5]
				\draw(0,0) pic[scale=1.5]{Uffolded};
			\end{tikzpicture}= (P\otimes P)(U \otimes_r\! U^*)^{\otimes_r 2}(P\otimes P).
			\label{eq:avegate}
		\end{align}
		Here $\otimes_r$ denotes the tensor product over replicas rather than spatial sites, and the operator
		\be
		P = \mathbb{E}[ (v\otimes_r v^*)^{\otimes_r 2}]\,\qquad v\in U(d),
		\ee
		is a projector on a $2-$dimensional space spanned by the vectors \eqref{eq:circrec} (see, e.g., Appendix G of Ref.~\cite{nahum2018operator} for an elementary proof). Note that these states are linearly independent but not orthogonal, indeed 
		\be
		\braket{\wrec}{\wcirc}=\frac{1}{d}\,.
		\label{eq:prod}
		\ee
		Since $P$ is a projector, we used   
		\be
		P^2=P,
		\ee
		to apply it also on the initial state matrix. Namely, we defined the averaged initial state matrix as  
		\begin{align}
			& n = \begin{tikzpicture}[baseline=(current  bounding  box.center),scale=1.5]	
				\path(0,0)pic[scale=1.5]{MPSpairstatef};
			\end{tikzpicture}=P (m\otimes_r m^*)^{\otimes_r 2} P. 
			\label{eq:avestate}
		\end{align} 
		
		The above discussion implies that all wires in \eqref{eq:Pt} carry a two-dimensional vector space spanned by $\ket{\wcirc}$ and $\ket{\wrec}$. Almost all matrix elements of $W$ and $n$ in this basis are fixed solely by dual unitarity \eqref{eq:dualunitarity} and the normalisation condition \eqref{eq:normalization}, and are hence independent of the specific $U$ and $m$. The only exceptions are 
		\begin{align}
			\begin{tikzpicture}[baseline=(current  bounding  box.center),scale=1]
				\path (0,0) pic[scale=2]{Uffolded};
				\draw[thick, fill=white] (0.5,1) pic[scale=1.5]{circ2};
				\draw[thick, fill=white] (-0.5,0) pic[scale=1.5]{circ2};
				\draw[thick, fill=white] (-0.5+0.1,1+0.1) rectangle (-0.5-0.1,1-0.1);
				\draw[thick, fill=white] (0.5+0.1,0.1) rectangle (0.5-0.1,-0.1);
			\end{tikzpicture}= 1-p+\frac{p}{d^2},\qquad 
			\scalebox{1}{\begin{tikzpicture}[baseline=(current  bounding  box.center),scale=1]	
					\draw (0,0) pic[scale=2]{MPSpairstatef};
					\draw[thick,fill=white] (-.5,1) pic[scale=2]{circ2};
					\draw[thick,fill=white] (.5,1) pic[scale=2]{rec2};
			\end{tikzpicture}}\equiv \frac{c}{d},
			\label{eq:defc}
		\end{align}
		where the second equation defines the parameter $c$, which characterises the averaged initial state matrix. In Appendix~\ref{sec:ranges} we show that $c$ takes values in $[1,d]$ and it is equal to one only when the initial state is solvable. 
		
		Considering for instance the orthonormal bases 
		\be
		\{\ket{\wcirc,\wcirc},\ket{\wcirc,\bcirc},\ket{\bcirc,\wcirc},\ket{\bcirc,\bcirc}\}, \quad \{\ket{\wcirc},\ket{\bcirc}\},
		\ee
		where we introduced the state 
		\begin{align}
			\ket{\bcirc} &=\frac{d\ket{\wrec}-\ket{\wcirc}}{\sqrt{d^2-1}},\label{eq:bcirc}
		\end{align}
		we explicitly find  
		\be
		W = \begin{bmatrix}1&0&0&0\\0&0&\textstyle 1-p& \frac{p}{\sqrt{d^2-1}}\\0&1-p&0&\frac{p}{\sqrt{d^2-1}}\\0&\frac{p}{\sqrt{d^2-1}}&\frac{p}{\sqrt{d^2-1}}&1-\frac{2p}{d^2-1}
		\end{bmatrix},
		\label{eq:avegateexpr}
		\ee
		and
		\be
		n = \begin{bmatrix}
			1&\frac{c-1}{\sqrt{d^2-1}}\\
			\frac{c-1}{\sqrt{d^2-1}}&1-\frac{2(c-1)}{{d}^2-1}\\
		\end{bmatrix}.
		\label{eq:explicitmave}
		\ee
		Since $W$ bares dependence on the gate only through $p$, the same holds for $\mathcal P_x$. We also stress that, since the averaged gate is symmetric and parity-invariant, we did not include a mark in its graphical representation. 	
		
		\subsection{Rigorous Proof of Property~\ref{prop:P1} for $p\geq \bar p(d)$}
		\label{sec:rigproof}
		
		In this subsection, we make use of the bound \eqref{eq:boundS2bar} to prove  Property~\ref{prop:P1} for $p > \bar p(d)$ (cf.~\eqref{eq:pbound}). To this end, we introduce the following lemma. 
		
		\begin{lemma}
			\label{lemma:LemmaPt}
			For $p\geq \bar p(d)$ and any state \eqref{eq:initialstate}, there exist  $A,B\geq 0$ such that 
			\be
			\begin{aligned}
				& d^x \mathcal P_x \leq {A + B x }.
			\end{aligned}
			\label{eq:asypurity}
			\ee
			The choice $B=0$ can only be made for initial \emph{solvable} states satisfying \eqref{eq:unitarym}. 
		\end{lemma}
		\noindent Eq.~\eqref{eq:asypurity} and \eqref{eq:boundS2bar} imply
		\be
		1 - \frac{\log(A + 2 B t)}{2 t \log d}\leq \frac{\bar S_A(t)}{4 t \log d} \leq 1.
		\ee
		Taking the infinite-time limit and using the bound \eqref{eq:boundvelocity}, we obtain
		\be
		v_{\rm E}=1,
		\ee
		which proves Property \ref{prop:P1}. Note that using Lemma~\ref{lemma:LemmaPt} one can also prove 
		\be
		v_{\rm E, \alpha}=1, \qquad \alpha \leq 2,
		\ee 
		by combining \eqref{eq:boundS2bar2}, \eqref{eq:nondec} and \eqref{eq:Sbound}.

		To prove Lemma~\ref{lemma:LemmaPt} we derive a simple recursive relation for $\mathcal P_x$. 
		We begin by noting that the dual-unitarity conditions \eqref{eq:dualunitarity} imply
		\begin{align}
			&\begin{tikzpicture}[baseline=(current  bounding  box.center),scale=1.5]
				\draw (0,0) pic[scale=1.5]{Uffolded};
				\draw  (-.25,.5) pic[scale=1.5]{circ2};	
				\draw  (.25,.5) pic[scale=1.5]{circ2};
			\end{tikzpicture}=
			\begin{tikzpicture}[baseline=(current  bounding  box.center),scale=1.5]
				\draw (-.25,.5)--(-.25,0);
				\draw (.25,.5)--(.25,0);
				\draw  (-.25,.5) pic[scale=1.5]{circ2};	
				\draw  (.25,.5) pic[scale=1.5]{circ2};
			\end{tikzpicture}\qquad\begin{tikzpicture}[baseline=(current  bounding  box.center),scale=1.5]
				\draw (0,0) pic[scale=1.5]{Uffolded};
				\draw  (.25,0) pic[scale=1.5]{circ2};
				\draw  (.25,0.5) pic[scale=1.5]{circ2};
			\end{tikzpicture}=\begin{tikzpicture}[baseline=(current  bounding  box.center),scale=1.5]
				\draw (.25,0)--(-.25,0);
				\draw (.25,.5)--(-.25,0.5);
				\draw  (.25,0) pic[scale=1.5]{circ2};
				\draw  (.25,0.5) pic[scale=1.5]{circ2};
			\end{tikzpicture},
			\label{eq:ducirc}
		\end{align}
		and 
		\begin{align}
			&\begin{tikzpicture}[baseline=(current  bounding  box.center),scale=1.5]
				\draw (0,0) pic[scale=1.5]{Uffolded};
				\draw  (-.25,.5) pic[scale=1.5]{rec2};	
				\draw  (.25,.5) pic[scale=1.5]{rec2};
			\end{tikzpicture}=\begin{tikzpicture}[baseline=(current  bounding  box.center),scale=1.5]
				\draw (-.25,.5)--(-.25,0);
				\draw (.25,.5)--(.25,0);
				\draw  (-.25,.5) pic[scale=1.5]{rec2};	
				\draw  (.25,.5) pic[scale=1.5]{rec2};
			\end{tikzpicture}\qquad\begin{tikzpicture}[baseline=(current  bounding  box.center),scale=1.5]
				\draw (0,0) pic[scale=1.5]{Uffolded};
				\draw  (.25,0) pic[scale=1.5]{rec2};
				\draw  (.25,0.5) pic[scale=1.5]{rec2};
			\end{tikzpicture}=\begin{tikzpicture}[baseline=(current  bounding  box.center),scale=1.5]
				\draw (.25,0)--(-.25,0);
				\draw (.25,.5)--(-.25,0.5);
				\draw  (.25,0) pic[scale=1.5]{rec2};
				\draw  (.25,0.5) pic[scale=1.5]{rec2};
			\end{tikzpicture}.
			\label{eq:durec}
		\end{align}
		Moreover, we have 
		\begin{align}
			\scalebox{1}{\begin{tikzpicture}[baseline=(current  bounding  box.center),scale=1]	
					\path(0,0)pic[scale=2]{MPSpairstatef};
					\draw (0.5,1) pic[scale=2]{circ2};
			\end{tikzpicture}}
			=\begin{tikzpicture}[baseline=(current  bounding  box.center),scale=1]	
				\path(0,0)pic[scale=2]{connection};
				\draw (0.5,1) pic[scale=2]{circ2};
			\end{tikzpicture}+\frac{c-1}{\sqrt{d^2-1}}\begin{tikzpicture}[baseline=(current  bounding  box.center),scale=1]	
				\path(0,0)pic[scale=2]{connection};
				\draw (0.5,1) pic[scale=2]{circ2b};
			\end{tikzpicture},\label{eq:staterel1}	\\\scalebox{1}{\begin{tikzpicture}[baseline=(current  bounding  box.center),scale=1]	
					\path(0,0)pic[scale=2]{MPSpairstatef};
					\draw (0.5,1) pic[scale=2]{rec2};
			\end{tikzpicture}}
			=\begin{tikzpicture}[baseline=(current  bounding  box.center),scale=1]	
				\path(0,0)pic[scale=2]{connection};
				\draw (0.5,1) pic[scale=2]{rec2};
			\end{tikzpicture}+\frac{c-1}{\sqrt{d^2-1}}\begin{tikzpicture}[baseline=(current  bounding  box.center),scale=1]	
				\path(0,0)pic[scale=2]{connection};
				\draw (0.5,1) pic[scale=2]{rec2b};
			\end{tikzpicture},\label{eq:staterel2}	
		\end{align}
		where we introduced 
		\be
		\ket{\brec}  =\frac{d\ket{\wcirc}-\ket{\wrec}}{\sqrt{d^2-1}}.
		\ee	
		We now have all the fundamental ingredients for deriving the desired recursive relations. Using \eqref{eq:staterel2} in the bottom right corner of \eqref{eq:Pt}, telescoping \eqref{eq:durec}, and using \eqref{eq:prod} we find
		\begin{align}
			\mathcal P_x= \frac{1}{d}\mathcal P_{x-1}+\frac{c-1}{\sqrt{d^2-1}}\mathcal Q_x,
			\label{eq:Pteq}
		\end{align}
		where we introduced~
		\be
		\mathcal Q_{x\geq2}= 
		\begin{tikzpicture}[baseline=(current  bounding  box.center),scale=1]	
			\foreach \i in {0,...,3}
			\path (\i ,0) pic{MPSpairstatef};
			\path (4,0) pic{connection};
			\foreach \i in {0,...,3}
			{\foreach \j in {0,...,\i}
				{\path (2-\i/2+\j ,2-\i/2) pic{Uffolded};
					\draw[thick, fill=white] (2-\i/2-0.25 ,2-\i/2+0.5) circle (0.07);}
				\draw[thick,fill=white] (2-\i/2+\i+0.25,2-\i/2+0.5) pic{rec2};
			}
			\foreach \i in {0,...,3}
			{
				\draw (2-\i/2+\i+0.25,2-\i/2+0.5) pic{rec2};
				\draw (2-\i/2-0.25 ,2-\i/2+0.5) pic{circ2};
			}	
			\foreach \i in {4}
			{			\draw[thick, fill=white] (2-\i/2-0.25 ,2-\i/2+0.5) pic{circ2};\draw[thick, fill=white] (2+\i/2+0.25 ,2-\i/2+0.5) pic{rec2b};
				\draw (2-\i/2+\i-0.125,-0.125+2-\i/2+0.5)--(2-\i/2+\i-0.25,2-\i/2+0.5);}
			\draw [decorate, decoration = {brace}]   (2.5,2.75) -- (4.75,0.5);
			\node[scale=1] 	at (4,1.7) {$x$};
		\end{tikzpicture},
		\label{eq:Qx}
		\ee
		and $\mathcal Q_{1}=\braket{\wcirc}{\brec}=\sqrt{d^2-1}/d$. Applying now \eqref{eq:staterel1} to the bottom left corner of \eqref{eq:Qx} and then telescoping \eqref{eq:ducirc}, we have 
		\begin{align}
			\mathcal Q_x=\frac{1}{d}\mathcal Q_{x-1}+\frac{c-1}{\sqrt{d^2-1}} \mathcal R_x,
			\label{eq:Qteq}
		\end{align}
		where 
		\be
		\mathcal R_x = \begin{tikzpicture}[baseline=(current  bounding  box.center),scale=1]	
			\foreach \i in {1,...,3}
			\path (\i ,0) pic{MPSpairstatef};
			\path (0,0) pic{connection};	\path (4,0) pic{connection};
			\foreach \i in {0,...,3}
			{\foreach \j in {0,...,\i}
				{\path (2-\i/2+\j ,2-\i/2) pic{Uffolded};
					\draw[thick, fill=white] (2-\i/2-0.25 ,2-\i/2+0.5) circle (0.07);}
				\draw[thick,fill=white] (2-\i/2+\i+0.25,2-\i/2+0.5) pic{rec2};
			}
			\foreach \i in {0,...,3}
			{
				\draw (2-\i/2+\i+0.25,2-\i/2+0.5) pic{rec2};
				\draw (2-\i/2-0.25 ,2-\i/2+0.5) pic{circ2};
			}	
			\foreach \i in {4}
			{			\draw (2-\i/2-0.25 ,2-\i/2+0.5) pic{circ2b};
				\draw[thick, fill=white] (2+\i/2+0.25 ,2-\i/2+0.5) pic{rec2b};		\draw (2-\i/2+\i-0.125,-0.125+2-\i/2+0.5)--(2-\i/2+\i-0.25,2-\i/2+0.5);}
		\end{tikzpicture}.
		\label{eq:Rt}
		\ee
		To close the recursive system formed by \eqref{eq:Pteq} and \eqref{eq:Qteq}, we now seek a bound for $\mathcal R_x$. In particular, a bound of the form   
		\be
		|\mathcal R_x| \leq \frac{C}{D^x},
		\label{eq:ineqR}
		\ee
		for some $C>0$ and $D>d$, leads to  
		\begin{align}
			\mathcal{Q}_x &\leq \frac{\alpha}{d^x}+\frac{\beta}{D^x},\\
			\mathcal{P}_x &\leq \frac{\gamma }{d^x}+  \frac{(c-1) \alpha}{\sqrt{d^2-1}}\frac{x}{d^x}+\frac{\delta}{D^x}, \qquad \alpha,\beta,\gamma,\delta \in \mathbb{R},
		\end{align}
		which immediately imply \eqref{eq:asypurity}.
		
		To find the bound in Eq.~\eqref{eq:ineqR}, we view $\mathcal R_x$ as the matrix element of 
		\be						
		[M_{1}]_{a;b} = \begin{tikzpicture}[baseline=(current  bounding  box.center),scale=1]	
			\path(-.5,0) pic{Uffolded};
			\path(.5,0) pic{Uffolded};
			\path(0,.5) pic{Uffolded};
			\path (0,-.5) pic{MPSpairstatef};
			\path (1,-.5) pic{MPSpairstatef};
			\path (-1,-.5) pic{MPSpairstatef};
			\path (1,-.5) pic{connection};
			\path (-1,-.5) pic{connection};
			\draw[line width=0.25mm] (-.15,1.75)--(.15,1.75);
			\node[scale=1] 	at (1.4-1,.2+1.035) {$b_{x-1}$};
			\node[scale=1] 	at (1.35-0.4,.2+0.4) {$\ddots$};
			\node[scale=1] 	at (1.35,.2) {$b_1$};
			\node[scale=1] 	at (.35,1.77) {$b_{x}$};
			\node[scale=1] 	at (-.35,1.75) {$a_{x}$};
			\node[scale=1] 	at (-1.3+1,.2+1) {$a_{x-1}$};
			\node[scale=1, rotate=90] 	at (-1.3+.4,.2+.4) {$\ddots$};
			\node[scale=1] 	at (-1.3,.2) {$a_1$};
		\end{tikzpicture},	
		\label{eq:Mmat}						
		\ee
		between the vectors 
		\be										
		[v_1]_b= \begin{tikzpicture}[baseline=(current  bounding  box.center),scale=1]	
			\foreach \i in {0,...,4}
			{\foreach \j in {0}															
				{
					\ifnum \i < 4
					\path (2+\i/2+\j ,2-\i/2) pic{Uffolded};
					\path (2+\i/2+\j +.25,2-\i/2+.5) pic{rec2};			\else 
					\path (2+\i/2+\j +.25,2-\i/2+.5) pic{rec2b};
					\path (2+\i/2+\j ,2-\i/2) pic{connection};
					\fi
					\path (2-.5+\j +.25,2+.5) pic{circ2};}	
			}
			\node[scale=1] 	at (3.15-1.5,.2+1.5) {$b_{x}$};
			\node[scale=1] 	at (3.15-.75,.2+.75) {$\ddots$};
			\node[scale=1] 	at (3.15,.2) {$b_1$};	
		\end{tikzpicture},	\label{diag:v}
		\ee							
		and 
		\begin{align}
			[w_1]_a=\begin{tikzpicture}[baseline=(current  bounding  box.center),scale=1]	\foreach \i in {0,...,3}
				{			\ifnum \i < 3
					\path (-\i/2,-\i/2) pic{Uffolded};
					\path (-\i/2-.25,-\i/2+.5)  pic{circ2};
					\else
					\path (-\i/2-.25,-\i/2+.5)  pic{circ2b};
					\path (-\i/2,-\i/2) pic{connection};
					\fi
				}
				\node[scale=1] 	at (-.5+1.1,-1.2+1.75) {$a_{x}$};
				\node[scale=1] 	at (-.5+1.2,-1.2+1) {$a_{x-1}$};
				\node[scale=1, rotate=90] 	at (-.5+.5,-1.2+.5) {$\ddots$};
				\node[scale=1] 	at (-.5,-1.2) {$a_1$};
			\end{tikzpicture}.
		\end{align}
		Employing the Cauchy-Schwartz inequality, we then obtain 
		\begin{align}
			|\mathcal R_x | = |\mel{v_1}{M_1}{w_1}| &\leq \|M_{1}\|_\infty \sqrt{\braket{v_1}{v_1}}\sqrt{\braket{w_1}{w_1}}. 
			\label{eq:CS}
		\end{align}
		Let us now consider separately the three factors on the r.h.s.. Since the gates are the average of dual-unitary gates, the operator norm of the dual averaged gate $\tilde{W}$, with elements 
		\begin{align}
			[\tilde{W}]_{(ab);(cd)} \equiv
			\begin{tikzpicture}[baseline=(current  bounding  box.center),scale=1.5]	
				\path (0,0.15) pic[scale=2]{Uffolded};
				\node at (-0.35,-0.2) {};
				\node at (-0.35,0.05) {$a$};
				\node at (0.35,0.05) {$c$};
				\node at (-0.35,1) {$b$};
				\node at (0.35,1) {$d$};
			\end{tikzpicture},
		\end{align}
		is one. Therefore, we have nontrivial contributions to the norm of $M_1$ only from the initial state row. This gives 
		\be
		\|M_{1}\|_\infty \leq  \|n\|_\infty^{x-2} = \left(\frac{d+c}{d+1}\right)^{x-2},
		\label{eq:Minfty}
		\ee
		where the identity  
		\be
		\|n\|_\infty =  \frac{d+c}{d+1},
		\ee
		is proven in Appendix~\ref{sec:ranges}. 
		
		Eq.~\eqref{eq:Minfty} is the \emph{key simplification} provided by dual-unitarity in the current setting. Even though the bottom boundary of the tensor network \eqref{eq:Mmat} is ``generic", the dual-unitarity of the bulk tensors implies that its operator norm is bounded by a number that scales with the number of tensors on its edge. This should be contrasted with the non-dual-unitary case, where the dual gate $\tilde{W}$ has operator norm greater than one~\cite{ippoliti2021postselectionfree}, and, therefore, the operator norm of $M_{1}$ scales with the total number of tensors composing it. In summary, because of dual-unitarity, Eq.~\eqref{eq:Minfty} shows an ``area scaling" rather than a ``volume scaling". As we now see, for large enough entangling power $p$ such an area scaling can be counter-balanced by the other two terms in \eqref{eq:CS}, leading to the bound \eqref{eq:ineqR}. 
		
		To treat the second factor in \eqref{eq:CS} we introduce the matrix 
		\be
		T_2=\begin{tikzpicture}[baseline=(current  bounding  box.center),scale=1.5,rotate around={0:(0,0)}]	
			\path (-0.2,-0.3525) pic[scale=1.5, rotate=-45]{Uffolded};
			\path (-1/2,0) pic[scale=1.5, rotate=135]{Uffolded};	
			\path (-1.1,-.17) pic[scale=1.5]{rec2};
			\path (.4,-.17) pic[scale=1.5]{rec2};
			\node at (0.2,-.2) {};
		\end{tikzpicture},
		\label{dia:T2}
		\ee
		so that we can write 
		\begin{align}
			\!\!\!\!{\braket{v_1}{v_1}} &=\! \begin{tikzpicture}[baseline=(current  bounding  box.center),scale=1.5,rotate around={0:(0,0)}]	
				\foreach \j in {0,...,3}
				{			\path (-0.2,-0.3525+\j * 0.7) pic[scale=1.5, rotate=-45]{Uffolded};
					\path (-1/2,\j * 0.7) pic[scale=1.5, rotate=135]{Uffolded};	
					\path (-1.1,-.17+\j * 0.7) pic[scale=1.5]{rec2};
					\path (.4,-.17+\j * 0.7) pic[scale=1.5]{rec2};
				}	\path (-0.028,-0.6) pic[scale=1.5]{rec2b};
				\path (-0.67,-0.6) pic[scale=1.5]{rec2b};
				\path (-0.028,2.3) pic[scale=1.5]{circ2};
				\path (-0.675,2.3) pic[scale=1.5]{circ2};
				\node at (0.2,-.2) {};
				\draw [decorate, decoration = {brace}]   (.6,2.1) -- (0.6,-0.4);
				\node at (.95,0.85) {$x-1$};
			\end{tikzpicture}\!= \mel{\wcirc\wcirc}{T^{x-1}_2}{\brec\brec},  
		\end{align}
		where we used the fact that the averaged gate is real. As shown in Appendix~\ref{app:transfermatrix}, the vector $\ket{\brec\brec}$ is an eigenvector of $T_2$ with eigenvalue 
		\be
		\label{eq:lambda}
		\lambda(p)=(1-p)^2+\frac{p^2}{d^2-1}.
		\ee
		Therefore, we have 
		\be
		{\braket{v_1}{v_1}}=  \lambda(p)^{x-1} \braket{\wcirc\wcirc}{\brec\brec} = \lambda(p)^{x-1}  \frac{d^2-1}{d^2}\,.
		\ee
		Proceeding analogously (cf. Appendix~\ref{app:transfermatrix}) we find  
		\be
		{\braket{w_1}{w_1}}=  \lambda(p)^{x-2}\,.
		\ee
		Finally, putting all together, we obtain the following bound  
		\begin{align}
			|\mathcal R_x | &\leq  \left(\frac{d+1}{d+c}\right)^2 \sqrt{\frac{d^2-1}{d^2 \lambda(p)^3}}  \left(\frac{d+c}{d+1}\right)^x\lambda(p)^x\label{eq:Retbound}\\
			&\leq  \left(\frac{d+1}{d+c}\right)^2 \sqrt{\frac{d^2-1}{d^2 \lambda(p)^3}}  \left(\frac{2d}{d+1}\right)^x\lambda(p)^x,
		\end{align}
		where we used that $c\leq d$. Choosing $p$ such that 
		\be
		\lambda(p) \frac{2d}{d+1} < \frac{1}{d},
		\label{eq:finineq}
		\ee
		we then find the bound \eqref{eq:ineqR}. Solving for $p$ we find that \eqref{eq:finineq} is indeed satisfied for all  
		\be
		p\geq \bar p(d)\,. 
		\ee
		This concludes the proof.

		\subsection{Extension to $p<\bar p(d)$}
		\label{sec:genericgates}
		
		An obvious strategy to generalise our proof is to extend Lemma~\ref{lemma:LemmaPt} to $p\leq \bar p(d)$. To this end, a simple observation is that, for small enough values of $c$, one can use the tighter bound \eqref{eq:Retbound} for $|\mathcal R_x |$. The latter grants the validity of Lemma~\ref{lemma:LemmaPt} whenever
		\be
		\left(\frac{d+c}{d+1}\right)\lambda(p)<\frac{1}{d}.
		\label{eq:general_bound}
		\ee
		Recalling that $c\geq 1$ (cf.~ Appendix~\ref{sec:ranges}) we find that this bound can be satisfied for some $c$ only if 
		\begin{align}
			p> \tilde p(d)\equiv \frac{d^2-1}{d^2}\left (1-\frac{1}{\sqrt{d+1}}\right).
		\end{align}
		In fact, the bound \eqref{eq:general_bound} can be easily refined. For instance, instead of Eq.~\eqref{eq:CS} we can consider
		\be
		\abs{\mathcal{R}_x}\le\sqrt{\braket{w_2}}\sqrt{\braket{v_2}}\norm{M_{2}},
		\ee
		with $\ket{w_2},\ket{v_2}$, defined with an extra row of gates, i.e. 
		\begin{align}
			v_2=\begin{tikzpicture}[baseline=(current  bounding  box.center),scale=1.25]		\foreach \i in {0,...,4}
				{\foreach \j in {0}															
					{
						\ifnum \i < 4
						\path (2+\i/2+\j ,2-\i/2) pic[scale=1.25]{Uffolded};
						\ifnum \i>0
						\path (1+\i/2+\j ,2-\i/2) pic[scale=1.25]{Uffolded};
						\fi
						\path (2+\i/2+\j +.25,2-\i/2+.5) pic[scale=1.25]{rec2};			
						\else 
						\path (2+\i/2+\j +.25,2-\i/2+.5) pic[scale=1.25]{rec2b};
						\path (2+\i/2+\j ,2-\i/2) pic[scale=1.25]{connection};
						\path (1+\i/2+\j ,2-\i/2) pic[scale=1.25]{MPSpairstatef};
						\fi
						\path (2-.5+\j +.25,2+.5) pic[scale=1.25]{circ2};	
						\path (1+\j +.25,2) pic[scale=1.25]{circ2};}
					\draw[thick,fill=myvioletc] (3,.25) circle (.07);
				}	
			\end{tikzpicture}.\label{eq:v2}
		\end{align}
		Comparing this with \eqref{diag:v}, we see that the norm $\braket{v_2}$ involves the matrix $T_4$. One can directly verify that the eigenvalue of $T_4$ contributing to this norm corresponds to an eigenvector with support $4$ and it is strictly smaller than $\lambda(p)$. This results in an immediate improvement of the bound. In fact, this procedure can be repeated considering increasingly ``thicker" states $\ket{w_x},\ket{v_x}$ for any $x\geq 2$ and leads to a systematic improvement. 
		\begin{figure}
			\includegraphics[width=\columnwidth]{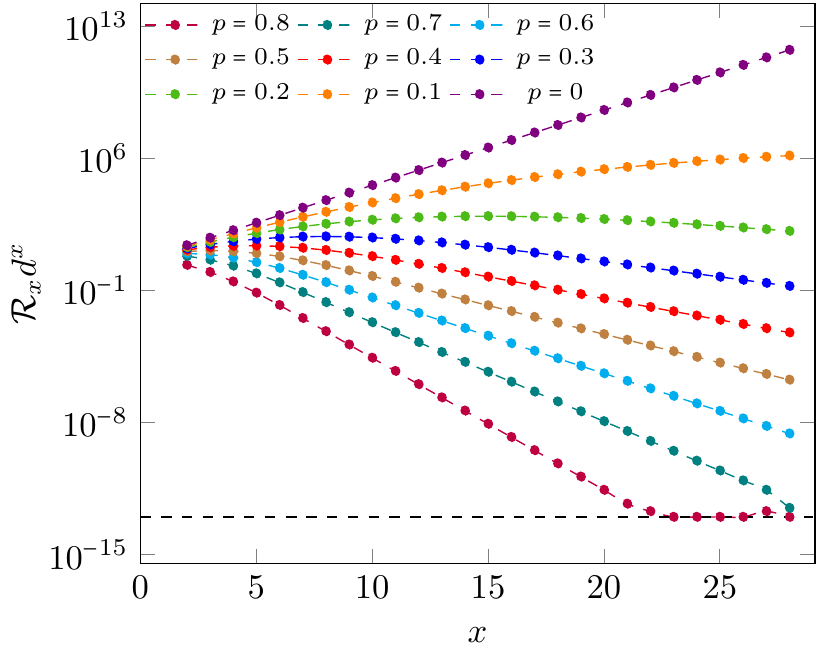}		
			\caption{Remainder $d^x \mathcal{R}_x$ (cf.~\eqref{eq:Rt}) for $c=2.5$, $d=5$, and several values of $p$. The exponential decay \eqref{eq:ineqR} with $D>d$ covers much more values than the range $p\gtrsim 0.62$,  for which our rigorous bound \eqref{eq:general_bound} applies. 
				The dotted line on the bottom indicates the limit of the numerical accuracy.}
			\label{fig:averaged_Rt_numerics}
		\end{figure}
		
		The fact that the bound on $\mathcal{R}_x$ can be improved is also suggested by numerical evidence. For instance, in Figure~\ref{fig:averaged_Rt_numerics} we show the behaviour of $\mathcal{R}_x d^x$ as a function of time for $d=5$. We see that the exponential decay \eqref{eq:ineqR} --- which implies the validity of Lemma~\ref{lemma:LemmaPt} --- is clearly shown by our numerical evaluations for $p>0.3$, which should be compared with $\bar p(5)\approx 0.68$ and $\tilde p(5)\approx 0.57$. From the trend in the numerical data, it is reasonable to expect that, upon accessing larger values of $x$, the same decay would be observed for all $p\neq 0$. A different indication is shown in Figure~\ref{fig:averaged_numerics}, which suggests that 
		\be
		\Delta \log \mathcal P_{x} \equiv {\log \mathcal P_{x-1}-\log \mathcal P_{x}},\label{def:deltaPt}
		\ee 
		approaches $1$ for all values of $p$ except for a neighbourhood of $p=0$. Consistently with Lemma~\ref{lemma:LemmaPt} the leading corrections at large $x$ appear to be $\approx x^{-1}$.

		\begin{figure}
			\includegraphics[width=\columnwidth]{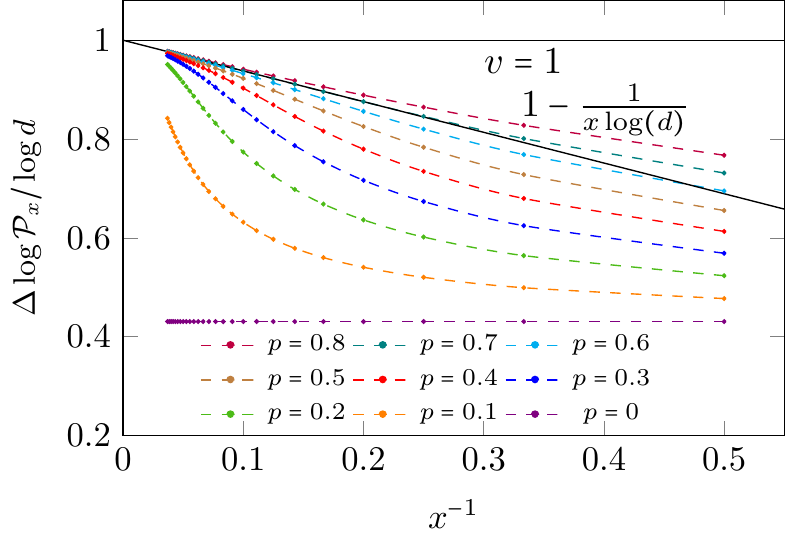}	
			\caption{Increment of  the indicator $\Delta\log(\mathcal{P}_x)$ in Eq.~\eqref{def:deltaPt} per step for an initial state corresponding to $c=2.5$ and $d=5$. The quantity is expected to saturate at the averaged velocity value as $x\ra\infty$. Note that our rigorous analytic bound \eqref{eq:general_bound} applies only if $p\gtrsim 0.62$.
				Assuming $d^x \mathcal{P}_x\sim {A+Bx}$ even for $p>0.62$, we expect, for large $x$, $\Delta \log \mathcal P_{x}  \sim 1- {1}/({x\log d})$.
			}
			
			\label{fig:averaged_numerics}
		\end{figure}
		
		\subsection{Numerical results for single realisations}
		\label{sec:numericsSR}
		
		In this subsection we provide numerical evidence supporting the claim that $v_{\rm E}$ is maximal for essentially \emph{any} single realisation of the gates \eqref{eq:uvmats}. For our numerical experiments we consider dual-unitary quantum circuits with a 2-dimensional local Hilbert space. In this case the most general local gate can be written as  
		\be
		U_{(ij);(kl)}= \!\!\!\!\sum_{i',j',k',l'=1}^2 \!\!\!\exp({\rm i} J \delta_{i'j'})u_{+\,i;i'}u_{-\,j;j'}v_{+\,k;k'}v_{-\,l;l'},
		\label{eq:Ud2}
		\ee
		where $\{v_{\pm},u_\pm\}$ are fixed $U(2)$ matrices and $J\in[0,\pi/2]$. The angle $J$ is in one-to-one correspondence with the entangling power~\cite{bertini2019exact}. Specifically, using the definition \eqref{eq:defp} we have  
		\be
		\label{eq:avsJ}
		p = \frac{2}{3} \cos(J)^2. 
		\ee
		In the following we use $p$, rather than $J$, to keep consistency with the previous subsections. The initial state matrix is instead taken of the form 
		\be
		m_{i;j}=(1-\delta_{ij})\sin(\theta)+\delta_{ij}\cos{\theta}\,.
		\label{eq:initialstatematrixd2}
		\ee
		Focussing on a space-time translationally invariant circuit, i.e., a circuit where the local gate is the same at each space-time point, we compute the R\'enyi entropy $S_A^{(2)}\!(t)$ for $t \leq L_A/2$ by numerically constructing the matrix $C_x$ (cf.~\eqref{diag:Ct}) and using \eqref{eq:RenyiCt}. This direct  approach allows us to reach values of $x$ up to $14$. Note that, due to the fast growth of entanglement in dual-unitary circuits, this is more efficient than tensor network methods based on the truncation of the time-evolving state, e.g., TEBD.  
		
		A representative example of our results is presented in Figure~\ref{fig:unaveraged_speed}, where we report $\Delta S_A^{(2)}(t)$ (cf. Eq.~\eqref{eq:entanglementincrement}) as a function of the inverse time. Our results suggest that at large times $\Delta S_A^{(2)}(t)$ approaches $4\log d$ with power law corrections that, as observed in the averaged case, are larger for smaller values of $p$. This implies 
		\be
		v_{\rm E,2}=v_{\rm E}=1,
		\ee 
		in accordance with our expectations. }	
	
	\begin{figure}
		\includegraphics[width=\columnwidth]{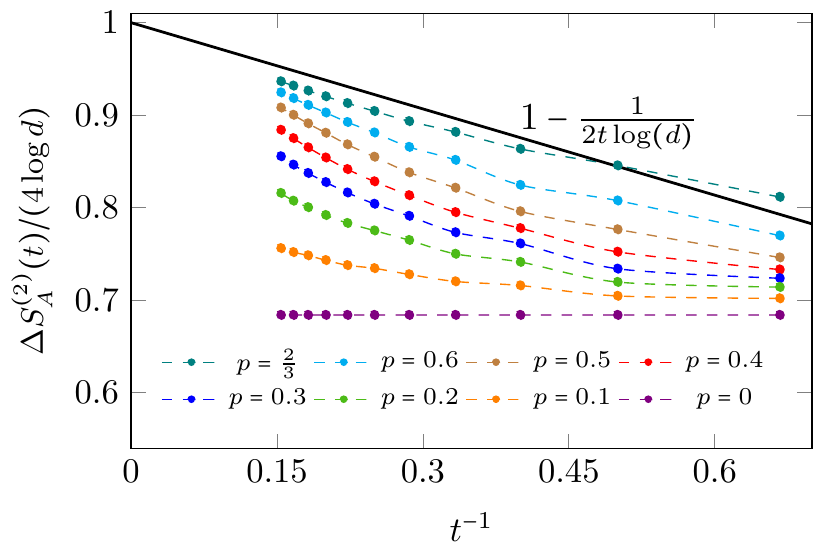}	
		\caption{$\Delta S_A^{(2)}(t)$ as a function of $t^{-1}$ for a system prepared in a generic pair product state specified by the matrices \eqref{eq:initialstatematrixd2} with $\theta=0.35$ and evolved with a homogeneous dual-unitary quantum circuit with local gate \eqref{eq:Ud2}, with fixed $\{v_{\pm},u_\pm\}$ (their explicit form is reported in Appendix~\ref{app:details_numerics}) and different values of $a$ (related to $J$ through \eqref{eq:avsJ}). This quantity saturates at the R\'enyi-2 entanglement velocity $v_{\rm E, 2}$ as $t\ra\infty$. Assuming that for large times $S_A^{(2)}(t)\sim -2 \log \mathcal P_{2t}$, we expect the various plots to reach asymptotically the line $1-{1}/({2t\log(d)})$ (black).}
		\label{fig:unaveraged_speed}
	\end{figure}
	
	\subsection{More general initial states}
	\label{sec:MPS}	
		
		In most of this paper we considered for simplicity the family of pair-product states~\eqref{eq:initialstate}. However, our approach can be applied more generally. For instance, as a non-trivial example let us consider a family of MPS that are generic fixed points of the RG flow~\cite{verstraete2005renormalization}, i.e., they can be written as
		\be
		\!\!\!\ket{\Psi_0}= \frac{1}{d^{\frac{L}{2}}} \sum_{a_j\in \mathbb Z_d} {\rm tr}[{Z^{{a}_1,{a}_2}}{Z^{a_3,a_4}}\ldots {Z^{a_{2N-1},a_{2L}}}] \ket{{a}},
		\label{eq:fixedpointMPS}
		\ee
		where $\{Z^{a,b}\}$ are $\chi\times\chi$ matrices such that 
		\be
		X\equiv \frac{1}{d}\sum_{a,b} Z^{a,b}\otimes (Z^{a,b})^*=\ketbra{L}{R},\label{eq:prop1MPS}
		\ee
		with 
		\be
		\ket{L}= \sum_{i=1}^{\chi} \ket{i,i}, \quad \ket{R}= \sum_{i=1}^{\chi} \lambda_i \ket{i,i}, \quad \sum_{i=1}^{\chi}  \lambda_i=1,\label{eq:prop2MPS}
		\ee
		and $\lambda_i>0$. Generalising the discussion of Secs.~\ref{sec:bound} and~\ref{sec:rigproof}, in Appendix~\ref{app:MPSinitialstate} we show that for  
		\begin{align}
			\lambda(p) \chi  \left(\frac{d+1}{d-1}+\sqrt{\frac{d+1}{d-1}}\right) <\frac{1}{d},\label{eq:conditionMPS}
		\end{align}
		Property~\ref{prop:P1} holds for all the MPS~\eqref{eq:fixedpointMPS}.

	\section{Discussion}%
	\label{sec:discussion}%
	
	We studied the asymptotic growth of entanglement in dual-unitary circuits prepared in generic low-entangled states. These states are generally \emph{non-solvable}: they break the unitarity of the evolution in space and their entanglement dynamics cannot be accessed using the standard dual-unitarity-based approaches~\cite{bertini2019entanglement, piroli2020exact}. Moreover, as opposed to solvable states, they display a sub-maximal entanglement increment at short times. 
	
	By introducing dual-unitarity-preserving random noise we showed that, surprisingly, the entanglement dynamics of generic states remain \emph{exactly tractable} for large times: one can still make exact statements for individual realisations of the noise, possibly excluding a subset with zero measure. In this way we proved that for a class of dual-unitary circuits with large enough entangling power the growth-rate of entanglement approaches the maximal value as time increases --- i.e.\ their entanglement velocity is always maximal irrespective of the initial conditions. We showed that this maximally entangling class exists for any number $d$ of local degrees of freedom as it includes the Hadamard family of dual-unitary gates introduced in Ref.~\cite{gutkin2020local}. Moreover, for $d\geq3$ it also contains 4-leg perfect tensors~\cite{facchi2008maximally, goyeneche2014genuinely, hosur2016chaos, huber2018bounds, rather2022thirty}. In fact, we presented analytical and numerical arguments suggesting that all dual-unitary circuits with non-zero entangling power belong to this class.
	
	Our results established an even tighter connection between dual-unitarity and maximal entanglement growth. While Ref.~\cite{zhou2022maximal} recently showed that if there exists an initial state for which the asymptotic entanglement rate is maximal, then the circuit is dual unitary, here we showed that in generic dual-unitary circuits \emph{every} initial state eventually approaches maximal entanglement growth. In this respect, our results show  that dual-unitary circuits are the hardest quantum circuits to simulate with classical computers~\cite{schuch2008entropy, schuch2008on, perales2008entanglement, hauke2012can} making of them the optimal test bed for investigations on quantum supremacy in the non-equilibrium dynamics~\cite{arute2019quantum}. 
	
	A natural question is whether our ``generality" assumption --- the fact that we excluded a zero-measure set of gates --- is necessary or not. Namely, do we need to exclude some special dual-unitary gates (e.g. the integrable ones) or \emph{any} dual-unitary circuit generates maximal entanglement growth at large times? 
	This would establish whether quantum chaos is an essential ingredient to produce the observed initial-condition independence of the entanglement velocities or dual unitarity alone suffices.

	Finally, we stress that the methods developed here provide a systematic way to investigate quenches from generic initial states in dual-unitary circuits. Interesting questions that one can tackle with them include (deep) thermalization timescales in dual-unitary circuits~\cite{bertini2019entanglement, piroli2020exact, claeys2022emergentquantum, ippoliti2022dynamical}, and multi-unital quantum channels~\cite{kos2022circuits}, or the ``temporal entanglement" scaling in chaotic quantum circuits~\cite{lerose2021influence} (see also Refs.~\cite{giudice2022temporal,lerose2021scaling,sonner2021influence,sonner2022characterizing}). The latter question is currently under investigation~\cite{foligno2022inprep}.

	\begin{acknowledgments}
		We thank Tianci Zhou for collaboration on a closely related project and for many useful comments on the manuscript. We also grateful to Lorenzo Piroli, Toma{\v z} Prosen, Pavel Kos, and, especially, Katja Klobas for very valuable feedback on the manuscript. This work has been supported by the Royal Society through the University
		Research Fellowship No.\ 201101.
	\end{acknowledgments}

	\appendix
	
	\section{Bounds on the entangling power of dual-unitary gates}
	\label{app:avegate}

	The values that the entangling power $p$ can take are bounded by the unitarity of the matrix.  To see this, consider  $\tr[({\tilde U}^{t_2} ({\tilde U}^{t_2})^\dag)^2]$ in the definition \eqref{eq:defp}:  the unitarity of the matrix fixes the value of  \begin{align}
		\tr[{\tilde U}^{t_2} ({\tilde U}^{t_2})^\dag]=\tr[{\tilde U} {\tilde U}^\dag]=d^2.
	\end{align}
	Applying \eqref{eq:boundpowertrace} to the matrix $\frac{{\tilde U}^{t_2} ({\tilde U}^{t_2})^\dag}{d^2}$,
	we find
	\begin{align}		
		\tr[({\tilde U}^{t_2} ({\tilde U}^{t_2})^\dag)^2]\in[d^2,d^4] \implies p\in[0,1].
	\end{align}
	In particular, the case 
	\be
	\tr[({\tilde U}^{t_2} ({\tilde U}^{t_2})^\dag)^2]=d^2
	\ee
	is attained if and only if ${\tilde U}^{t_2}$ is unitary, having all eigenvalues with magnitude $1$. This request, together with the dual unitarity conditions~\eqref{eq:dualunitarity}, means that $U$ is unitary for any choice of couples of in/out indexes. Tensors with this property are known as 4-leg perfect tensors and they exist for all $d>2$~\cite{huber2018bounds, rather2022thirty}. There is, however, a non-exhaustive class of dual unitary gates which is well defined in any dimension~\cite{claeys2021ergodic}
	\begin{align}
		U_{(ij),(kl)}=\delta_{il}\delta_{jk} \exp({\rm i} J_{ij}),
		\label{eq:LC}
	\end{align}
	with $J_{ij}$ being any set of $d^2$ real numbers.
	In terms of $J_{ij}$, we can write
	\begin{align}
		\!\!\tr[({\tilde U}^{t_2} ({\tilde U}^{t_2})^\dag)^2]\!=\!\!\!\!\!\sum_{i,j,k,l} \!\!\!\!{\exp[{\rm i}(J_{ij}+J_{kl}-J_{il}-J_{kj})]}\,.
	\end{align}
	As before, we can express the right-hand side as a matrix trace
	\begin{align}
		\!\!\!\!\sum_{i,j,k,l} {\exp[{\rm i}(J_{ij}+J_{kl}-J_{il}-J_{kj})]}=d^4\,\tr[(\xi\xi\dagg)^2] ,
	\end{align}
	with 
	\be
	\xi_{ij} \equiv \frac{\exp({\rm i} J_{ij})}{d}\quad \implies \quad {\Tr[\xi \xi^{\dag}]}=1. 
	\ee
	Using again \eqref{eq:boundpowertrace}, we find
	\begin{align}
		\tr[({\tilde U}^{t_2} ({\tilde U}^{t_2})^\dag)^2]	\in\left[d^3,d^4\right]	\,\,\implies\,\,  p\in \left[0,\frac{d}{d+1}\right].
	\end{align}
	The choice $J_{ij}=0$, which corresponds to a swap gate, gives $p=0$. Instead, the choice 
	\be
	J_{ij}=\frac{2\pi ij}{d},
	\ee
	corresponding to the Hadamard family	considered in Section~\ref{sec:randomDU}, gives 
	\be
	p=\frac{d}{d+1}.
	\ee
	We also note that, since the value of $p$ depends continuously on $J_{ij}$, there must exist gates corresponding to all values in the range
	\be
	p\in\left[0,\frac{d}{d+1}\right].
	\ee 
	Finally, we remark that this range is exhaustive in $d=2$, since any dual unitary gate can be expressed as in Eq.~\eqref{eq:Ud2}.

	\section{Averaged initial state matrix}
	\label{sec:ranges}
	Considering the averaged form of the initial state matrix in the basis $\{\ket{\wcirc},\ket{\bcirc}\}$ we have 
	\begin{align}
		n =\begin{pmatrix}
			1&\frac{c-1}{\sqrt{d^2-1}}\\
			\frac{c-1}{\sqrt{d^2-1}}&1-2 \frac{c-1}{{d}^2-1}\\
		\end{pmatrix},
		\label{eq:explicitmave}
	\end{align}
	with
	\be
	c = \frac{1}{d} \tr((m\dagg m)^2).
	\ee
	We can bound the values that the constant $c$ can take noting that the matrix $m$ is subject to the constraint
	\be
	\tr(m\dagg m)=d. 
	\ee
	Therefore, we can use \eqref{eq:boundpowertrace} on ${m\dagg m^{\phantom{\dag}}}/{d}$ with $\mathcal N=d$, $\alpha=2$, finding
	\be
	c\in[1,d].
	\ee
	The matrix \eqref{eq:explicitmave} is Hermitian and, therefore, its operator norm coincides with the norm of its maximal eigenvalue. Computing it explicitly, we find
	\begin{align}
		\lambda_{max}&=1-\frac{c-1}{d^2-1}+\sqrt{\left(\frac{c-1}{\sqrt{d^2-1}}\right)^2+\left(\frac{c-1}{d^2-1}\right)^2}\notag\\
		&=1-\frac{c-1}{d^2-1}\left(1-\sqrt{d^2-1+1}\right)=\frac{d+c}{d+1}. 
	\end{align}
	In summary we have 
	\begin{align}
		\|n\|_{\infty} = \frac{c+d}{{d+1}}\in \left [1,\frac{2d}{d+1}\right ]\,.
	\end{align}
		
		\section{Initial states in MPS form}
		\label{app:MPSinitialstate}
		
		Consider an MPS which is a generic fixed point  of the RG flow, i.e. it obeys Eqs \eqref{eq:prop1MPS}, \eqref{eq:prop2MPS}. This means we can write 
		\begin{align}
			X\dagg X={\braket{L}}   {\ketbra{R}}
		\end{align} 
		implying that 
		\begin{align}
			\norm{X}_\infty=\sqrt{{\braket{R}}}\sqrt{{\braket{L}}}\le\sqrt{\chi}.\label{eq:Xnorm}
		\end{align}
		We can repeat the steps of Sec \ref{sec:rigproof} finding recurrence relations for $\mathcal{P}^{MPS}_x,\mathcal{Q}^{MPS}_x$ --- the analogues of \eqref{eq:Pt} and \eqref{eq:Qx} --- and bounding $\mathcal{R}^{MPS}_x$ --- the analogue of \eqref{eq:Rt} --- by means of the Cauchy-Schwartz inequality. In order to do so, we find the operator norm of the matrix representing the folded MPS after averaging over the local unitaries. 
		
		We begin by noting that the average restricts the upper indices $(a_1,a_2,a_3,a_4)$ and $(b_1,b_2,b_3,b_4)$ of the doubly folded tensor  
\be
Z^{a_1,b_1}\otimes (Z^{a_2,b_2})^*\otimes Z^{a_3,b_3}\otimes (Z^{a_4,b_4})^*,
\ee
to the subspace spanned by  
\be
\ket{\wcirc}=\frac{\delta_{a_1,a_2}\delta_{a_3,a_4}}{d},
\label{eq:wcirc}
\ee
and 
\be
\ket{\wrec}=\frac{\delta_{a_1,a_4}\delta_{a_3,a_2}}{d}.
\label{eq:wrec}
\ee
We call $\widetilde{Z}$ the resulting tensor. 

Grouping the indices $(a,i)$ and $(b,j)$, the tensor $\widetilde{Z}^{a,b}_{i,j}$ can be seen as a matrix of dimension $2\chi \times 2\chi$, acting on the vector space $V_{\wcirc,\bcirc}\otimes V_{\chi}$, where $V_{\wcirc \bcirc}$ is the space spanned by \eqref{eq:wcirc} and \eqref{eq:wrec}, and $V_{\chi}= \mathbb C^\chi$ is the auxiliary space of the MPS. Our task is to find the operator norm of this matrix. To this end, we write it as		
\begin{align}
\widetilde{Z}=\begin{pmatrix}
\widetilde{Z}^{\wcirc\wcirc}& \widetilde{Z}^{\wcirc\bcirc}\\
\widetilde{Z}^{\bcirc\wcirc}& \widetilde{Z}^{\bcirc\bcirc}
\end{pmatrix},
\end{align}
and bound its norm as
\begin{align}
\norm*{\widetilde{Z}}_{\infty}\le \max\left\{\norm*{\widetilde{Z}^{\wcirc\wcirc}}_\infty,\norm*{\widetilde{Z}^{\bcirc\bcirc}}_\infty\right\}+\norm*{\widetilde{Z}^{\wcirc\bcirc}}_\infty. 
\label{eq:triangineq}
\end{align}
Using Eqs.~\eqref{eq:prop1MPS} and \eqref{eq:Xnorm} we find
\begin{align}
\norm*{\widetilde{Z}^{\wcirc\wcirc}}_\infty=\norm*{\widetilde{Z}^{\wrec\wrec}}_\infty=\left(\norm*{X}_\infty\right)^2\le\chi.
\end{align}
Consider now the $n$-th norm of $\widetilde{Z}^{\wcirc\wrec}$ (where $n$ is taken to be even): we can write it as 
\begin{align}
\norm*{\widetilde{Z}^{\wcirc\wrec}}_n	= \left({\rm Tr}(A_n A_n\dagg A_n A_n\dagg)\right)^{\frac{1}{n}},
\end{align}
where we defined the matrix
\be
\left(A_n\right)_{{a},{b}} \equiv d^{-\frac{n}{2}} {\rm Tr}\left[\prod_{i=1}^{{n}/{2}} Z^{a_{2i-1},b_{2i-1}}\left(Z^{a_{2i},b_{2i}}\right)^\dagger\right].
\label{eq:Amatdef}
\ee		
Here $a_i=1,\ldots d$ and the trace is taken over the auxiliary space.  Moreover, we have that
\begin{align}
{\rm Tr}(A_n A_n\dagg)={\rm Tr}\left[{\left(X X\dagg\right)}^\frac{n}{2}\right].
\end{align}
The matrix $A_n A_n\dagg$ is Hermitian and acts on a $d^n-$dimensional vector space, this means that
\begin{align}
\frac{{\rm Tr}(A_n A\dagg_nA_n A\dagg_n)}{{\rm Tr}\left[\left(X X\dagg\right)^\frac{n}{2}\right]^2}\in[d^{-n},1],
\end{align}
Taking the limit $n\ra \infty$, and using Eq.~\eqref{eq:Xnorm}, we find
\begin{align}
\!\!\norm*{\widetilde{Z}^{\wcirc,\wrec}}_\infty\in \left[\frac{\norm{X}_\infty^2}{d},\norm{X}_\infty^2\right]\implies 		\norm*{\widetilde{Z}^{\wcirc,\wrec}}_\infty<\chi.
\label{eq:normcircrec}
\end{align}
Using Eqs.~\eqref{eq:bcirc}, \eqref{eq:normcircrec} and the triangular inequality, the norms of the submatrices are bounded as 
\begin{align}
\norm*{\widetilde{Z}^{\wcirc\bcirc}}_\infty\le\chi \sqrt{\frac{d+1}{d-1}}\qquad \norm*{\widetilde{Z}^{\bcirc\bcirc}}_\infty\le \chi {\frac{d+1}{d-1}},
\end{align}
finally, Eq.~\eqref{eq:triangineq} implies
\begin{align}
\norm{\widetilde{Z}}_\infty\le\chi\left( \frac{d+1}{d-1}+\sqrt{\frac{d+1}{d-1}}\right).
\end{align}
Using this result, we finally obtain condition \eqref{eq:conditionMPS}.

\section{Two-site transfer matrix}
\label{app:transfermatrix}
Consider the transfer matrix $T_2$ in Eq.~\eqref{dia:T2}: using the explicit expression for the averaged gate \eqref{eq:avegateexpr} (which has the same form in the square states basis), we see that its explicit form in the basis
	\be
	\{\ket{\wrec\wrec},\ket{\brec\wrec},\ket{\wrec \brec},\ket{\brec \brec}\},
	\ee
	reads as
	\begin{align}
		T_2=\begin{pmatrix}
			1&0&0&0\\
			0&1-p&0&0\\
			0&0&1-p&0\\
			0&0&0&(1-p)^2+\frac{\displaystyle p^2}{\displaystyle{d^2}-1}
		\end{pmatrix},
	\end{align}
	allowing us to immediately compute
	\begin{align}
		{\braket{v_2}{v_2}} &=  \mel{\wcirc\wcirc}{T^{x-1}_2}{\brec\brec}\notag\\
		&=\lambda(p)^{x-1}\braket{\wcirc\wcirc}{\brec\brec}=\notag\\
		&=\lambda(p)^{x-1}\frac{d^2-1}{d^2}. 
	\end{align}
	Similarly, the transfer matrix
	\begin{align}
		T_2'=	\begin{tikzpicture}[baseline=(current  bounding  box.center),scale=1.5,rotate around={0:(0,0)}]	
			\foreach \i in {0,...,1}
			{			
				\path (-\i/2,0) pic[scale=1.5, rotate=-45]{Uffolded};
			}
			\path (.35-1.05,0.175) pic[scale=1.5]{circ2};
			\path (.35+.2,0.175) pic[scale=1.5]{circ2};
			\node at (0.2,-.2) {};
		\end{tikzpicture},
	\end{align}
	can be put in the same form under a unitary change of basis  $\{\wrec,\brec\} \ra \{\wcirc,\bcirc\}$, allowing us to find\begin{align}
		{\braket{w_2}{w_2}} &=  \mel{\bcirc \bcirc}{{T'}^{x-2}_2}{\,\arcstate}\notag\\
		&=\lambda(p)^{x-2}\braket{\bcirc \bcirc}{\,\arcstate}\notag\\
		&=\lambda(p)^{x-2},
	\end{align}
	where we used the notation for the arc state 
	\be
	\ket*{\,\arcstate}=\ket{\wcirc\wcirc}+\ket{\bcirc\bcirc}.
	\ee

	\section{Details of the numerics}\label{app:details_numerics}
	In Fig.~\ref{fig:unaveraged_speed} we plotted the numerical evaluation of the R\'enyi 2 entropy for different values of the parameter $p$.
	We implemented the gates defined in Eq.~\eqref{eq:Ud2}, where the unitary matrices $u_\pm,v\pm\in U(2)$ are obtained from a random Haar uniform extraction. Those matrices are kept fixed while varying the value of $J$ (or $p$, which are connected through Eq.~\eqref{eq:avsJ}). The explicit parameterisation implemented is the following
	\begin{align}
		\!\!\!\begin{bmatrix}
			\cos(\alpha)+{\rm i} \sin(\alpha)\cos(\theta)& {\rm i} \sin(\alpha)\sin(\theta) e^{-{\rm i}\phi}\\ {\rm i} \sin(\alpha)\sin(\theta) e^{{\rm i}\phi}&
			\cos(\alpha) - {\rm i} \sin(\alpha)\cos(\theta)
		\end{bmatrix}.
	\end{align}
	The values  used to produce Fig.~\ref{fig:unaveraged_speed} are reported in Tab.~\ref{table:parameters_unitary}.
	\begin{table}[b]
		\begin{tabular}{|c|c|c|c|}
			\hline
			& $\theta$ & $\phi$ & $\alpha$ \\
			\hline
			$u_-$ & 0.774764 & 5.531527 & 4.534001 \\
			\hline
			$u_+$ & 2.521203 & 3.352128 & 4.712387  \\
			\hline
			$v_-$ & 1.768693 & 0.704289 & 5.567499 \\
			\hline
			$v_+$ & 0.251880 & 1.607363 & 5.823117 \\
			\hline
		\end{tabular}
		\caption{Parameters for the one-site unitaries used to produce the data in Figure \ref{fig:unaveraged_speed}.}
		\label{table:parameters_unitary}
	\end{table}

	\bibliography{bibliography2.bib} 
	
\end{document}